\documentclass[pdflatex,sn-basic, authoryear]{sn-jnl}

\usepackage{graphicx}%
\usepackage{multirow}%
\usepackage{amsmath,amssymb,amsfonts}%
\usepackage{amsthm}%
\usepackage[title]{appendix}%
\usepackage{xcolor}%
\usepackage{textcomp}%
\usepackage{manyfoot}%
\usepackage{booktabs}%
\usepackage{algorithm}%
\usepackage{algorithmicx}%
\usepackage{algpseudocode}%
\usepackage{listings}%

\usepackage[scr=rsfso]{mathalfa}

\unnumbered 

\begin{document}

\title[Article Title]{Dynamic Functional Connectivity Resolves Brain Integration–Segregation Trade-off Under Costly Links}


\author*[1]{\fnm{Simachew Abebe} \sur{Mengiste}}\email{mengiste@unistra.fr}

\author*[1]{\fnm{Demian} \sur{Battaglia}}\email{dbattaglia@unistra.fr}

\affil*[1]{\orgdiv{Laboratoire de Neurosciences Cognitives et Adaptatives - CNRS UMR 7364}, \orgname{Universit\'e de Strasbourg}, \orgaddress{\street{12 rue Goethe}, \city{Strasbourg}, \postcode{67000}, \country{France}}}


\abstract{Dynamic functional connectivity (dFC) is ubiquitously observed in the brain, but why functional networks should remain dynamic even at rest is unclear. We asked whether temporal reconfiguration becomes advantageous when keeping a functional link active is costly. Modeling resting-state dFC as a temporal communication network, we show that empirical dFC outperforms equal-cost static architectures by increasing the reach and speed of information spreading in sparse regimes. Unlike more randomized temporal null models, however, it also preserves strong local cohesiveness, temporal clustering, rapid return of information to its source, and high neighborhood retention. Empirical dFC therefore achieves a compromise between large-scale integration and transient local segregation. This compromise is not explained by generic temporal variability, nor by partially frozen null models with persistent templates. A connectome-based mean-field model reproduces several key features, including high spatial and temporal clustering and strong integrative and segregative performance, but remains more stable over time than the empirical data. Our results indicate that empirical dFC reflects a structured regime of controlled persistence and renewal, in which local neighborhoods are maintained long enough to support transient recirculation before broader network-wide spreading occurs. Dynamic functional connectivity thus appears to be a resource-efficient solution to competing communication demands.}


\maketitle

\section{Introduction}

Functional connectivity ---describing communication and interaction between brain regions or neurons--- is intrinsically dynamic. Across imaging modalities and analysis frameworks, functional network links fluctuate over time, not only during changes in task, behavior, or context, but also during unconstrained rest \citep{Hutchison2013, Calhoun2014, Preti2017}. Such observations from whole-brain neuroimaging extend as well to finer microcircuit scales, where transiently synchronous neuronal assemblies continuously form and dissolve, giving rise to functional networks with shifting core–periphery and temporal rich club organization rather than fixed interaction patterns \citep{Pedreschi2020, Pedreschi2022}. Electrophysiological recordings and brain imaging further indicate that even healthy baseline activity is better described as a structured spatiotemporal flow of network reconfigurations than as a stationary architecture for information exchange \citep{Arbabyazd2023, Clawson2023, Pedreschi2026}. Across scales and conditions, therefore, dynamic functional connectivity (dFC) appears not as an exception, but as a pervasive feature of brain organization.

What remains unclear is why this dynamicity is so ubiquitous. One possibility is that dFC is merely a mechanistic consequence of the dynamical regimes in which neural systems operate, displaying metastability, multistability, or other forms of complexity \citep{Deco2013, Hansen2015, Cabral2017, Pathak2025}. Another possibility is that dFC is not directly functional, but instead reflects latent internal computations or state transitions that are functionally relevant, accounting for the many associations reported between structured features of dFC and cognition, aging, or pathology \citep{Bassett2011, Braun2015, Shine2016, Battaglia2020, Lombardo2020}.

Here we investigate a stronger possibility: that the dynamicity of functional connectivity is itself advantageous. We test the idea that, when sustaining functional links carries a significant cost, dynamic architectures can outperform static ones for communication among neural populations. Under this hypothesis, dFC is not a mere by-product of brain dynamics, but a resource-efficient strategy: by temporally reconfiguring which links are active, a system can reuse a limited interaction budget across time instead of maintaining all links simultaneously. This may allow the dynamic network to enable richer patterns of communication while using fewer resources.

Such a mechanism is especially relevant to a central organizational constraint of brain function: the coexistence of integration and segregation. Efficient cognition requires distributed populations to interact and combine information, but also to preserve partial autonomy, specialization, and diversification of activity \citep{Tononi1994, Sporns2013, Deco2015, Shine2019}. Maximizing integration alone tends to collapse distinct subsystems into overly coherent collective dynamics, whereas maximizing segregation alone limits large-scale coordination and information sharing. Brain function therefore depends on a balance between these competing demands. We hypothesize that dFC can help resolve this trade-off under costly links, by allowing patterns of integration to be deployed sequentially in time while preserving segregation at any given moment.

In this study, we formalize and test this principle in a minimal framework. We ask whether temporal reconfiguration of functional connectivity can increase the coexistence of integration and segregation relative to more static regimes under a comparable link cost, and whether this advantage provides a principled explanation for the pervasive dynamicity of brain functional connectivity. More broadly, our aim is to examine whether the restless character of functional connectivity reflects a fundamental design principle of neural communication.

Our results show that when establishing and maintaining functional links over time is costly ---e.g. energetically or in terms of limited synaptic resources \citep{Raichle2006, Tomasi2013, Volpi2024}---, dynamic connectivity provides a clear communication advantage. With only a limited number of links, dynamically reconfiguring them allows information to reach more regions and to do so faster on average than in more static architectures with comparable sparsity and equal cost. At the same time, empirical connectomes do not appear optimized for integration alone. Rather, they support a subtler compromise in which information can first circulate within local neighborhoods before being dispersed more broadly, thereby preserving segregation while still enabling large-scale diffusion. Strikingly, neither temporal network null models \citep{Kivela2012, Gauvin2013} nor mechanistic connectome-based mean-field models \citep{Hansen2015, Pathak2025} reproduced this joint integrative and segregative performance as displayed by empirically observed resting-state dynamic functional connectivity networks, suggesting that non-trivial optimization processes shape their spatiotemporal organization.

In summary, under costly links, dynamic functional connectivity is not a drawback or a contingent by-product of brain activity ---it could be a resource-efficient solution to the competing demands of brain communication.

\setcounter{figure}{0}

\section{Results}

\subsection{Information dispatch on cost-constrained dynamic functional connectivity}

\begin{figure}[H]
\centering
\includegraphics[width=\textwidth]{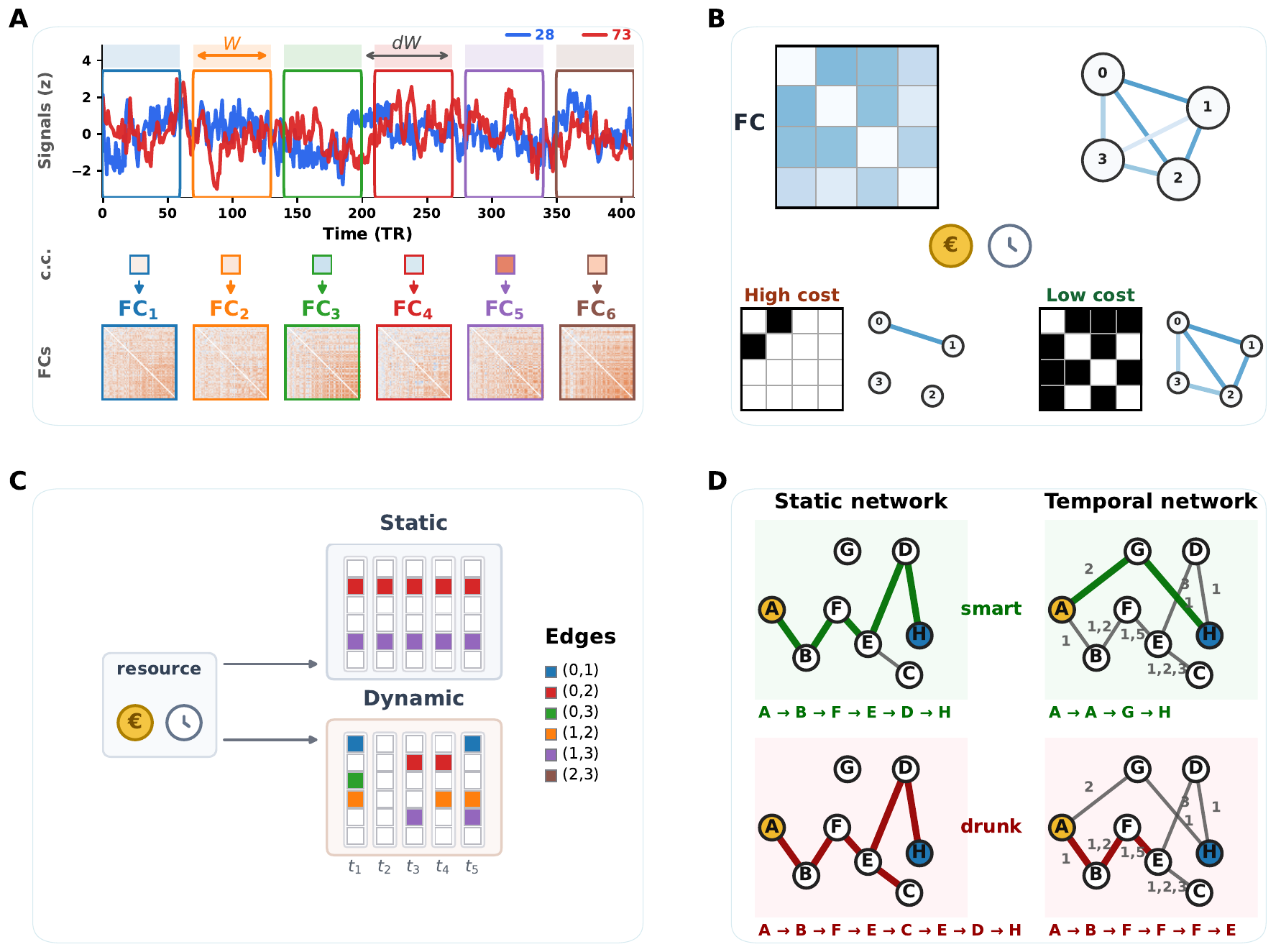}

\caption{\textbf{Information dispatch on dynamic functional
connectivity.} A, Static functional connectivity (FC) is estimated as the correlation structure of regional activity time series over long recording periods (for example, a 30-min resting-state fMRI session). To resolve dynamic functional connectivity (dFC) and generate a time-ordered stream of functional networks, a sliding-window approach can be used: successive FC frames, FC$_1$, FC$_2$, …, FC$_t$, are computed over shorter windows of duration $W$ (here 20 TR, i.e. $\approx 14$\,s), shifted by a step $dW$ (here 1 TR, i.e. 720\,ms). B, The resulting weighted networks are then binarized to retain only a subset of links compatible with a prescribed budget. When the cost of maintaining a link active for one time step is high, the resulting frames are sparser and have lower mean degree (left). Lower activation costs instead yield denser network frames (right). C, For a fixed unitary link cost, the same total budget can be allocated either to fewer, longer-lasting links (top) or to more numerous, shorter-lived and more transient links (bottom). In the schematic representation, each colored pixel denotes the activation of a given link (row) at a given time step (column). D, Once the temporal network has been fully constructed under the budget constraint, communication is modeled as the dispatch of discrete information packets that can either hop to a connected node or wait at the current node at each time step. A packet can move from node $A$ to node $B$ at time $t$ only if the link between $A$ and $B$ is active at that time. Here, numbers on links indicate the time steps at which each link is active; for example, label (1,5) denotes a link active in frames 1 and 5. We compare communication on equal-cost static and temporal networks (left and right columns, respectively) for two dispatch strategies: smart and drunk (top and bottom rows, respectively). The sparser static network corresponds to the time-1 frame of the example temporal network. Smart dispatch follows the fastest available route from a source (here node $A$) to a target (here node $H$), whereas drunk dispatch moves randomly along available links, including self-loops corresponding to waiting. In static networks, the fastest route coincides with the shortest path. In temporal networks, however, faster routes may emerge through time-dependent shortcuts. In the example shown, the smart walker on the static network reaches $H$ in five hops, whereas on the temporal network it first waits at node $A$ and then exploits links appearing at successive time steps to follow a shorter time-respecting route. Dynamic reconfiguration is not always advantageous: in the bottom-right example, the drunk walker reaches node $F$ via a link active at time 2, but then remains trapped because no outgoing link becomes available before time 5. }\label{fig1}
\end{figure}

Dynamic functional connectivity (dFC) is commonly estimated by computing functional connectivity ---zero-lag Pearson correlation between time-series or other more elaborate metrics--- over successive temporal windows of neuroimaging time series, thus generating a sequence of time-resolved network frames rather than a single static graph \citep{Hutchison2013, Preti2017, Battaglia2020}. In graph-theoretical terms, this sequence defines a \emph{temporal network}, or equivalently a particular form of multiplex network in which each layer corresponds to the connectivity observed in one temporal frame and layers are ordered in time \citep{Holme2012, Kivela2014}. In the empirical data considered here, each frame is initially a weighted connectivity matrix derived from sliding-window correlations of BOLD activity (Fig.~\ref{fig1}A). We begin by binarizing these weighted frames, so that links are either present or absent at each time step. This simplification deliberately sets aside variations in link strength and yields the minimal setting in which to study how the temporal allocation of functional interactions shapes communication possibilities.

The rationale for viewing functional connectivity as a communication network comes from information theory and from accumulating empirical evidence that functional couplings constrain the routes along which perturbations and information can spread. Measures such as Pearson correlation capture the linear component of statistical dependence between regional signals and can be interpreted, at least abstractly, as indicators that a communication channel is effectively open between two nodes at a given time \citep{Shannon1948}. This does not imply that information is necessarily exchanged through a direct anatomical projection, nor that functional connectivity should be identified with structural wiring. Rather, the presence of a functional link indicates that the activity of two nodes is coupled in a way compatible with above-chance information sharing, whether this arises through direct transmission, polysynaptic interactions, shared drive, or collective synchronization. Consistent with this view, experimental and modeling studies have shown that the propagation of perturbations is often better predicted by functional than by structural connectivity, indicating that functional links act as dynamically expressed conduits for communication \citep{Grothe2018, Papadopoulos2020}.

On this basis, we introduce a toy model of information dispatch on dFC networks. The aim is not to provide a biophysically complete account of neural communication, but to isolate one abstract question: if functional links define momentary communication opportunities, how should a limited budget of such opportunities be distributed across time to maximize information spreading? In this toy model, an information packet can traverse one active link per time step, and only if that link is present at that moment. Time is discretized, network links may be reconfigured from one frame to the next, and packets are propagated on the temporal functional network itself rather than on the underlying structural connectome. This abstraction lets us compare communication performance across alternative temporal organizations while keeping the underlying notion of “communication opportunity” as simple as possible.

Binarization also makes it possible to formalize the notion of link cost. Applying different thresholds to weighted dFC frames generates temporal networks with different instantaneous densities: stricter thresholds retain fewer links per frame (higher link cost regimes) and therefore correspond to sparser network frames, whereas looser thresholds retain more links (lower link cost regimes) and correspond to denser frames (Fig.~\ref{fig1}B). We assume that the relevant resource is spent on the presence of a link during a time step, regardless of whether that link has just appeared or has persisted from the previous frame. Under this assumption, the total cost of a temporal network over an observation window is the total number of active links per frame summed across all frames. The same overall budget can therefore be allocated either to many short-lived links that reconfigure rapidly over time or to fewer links that persist longer. This defines the central comparison of the present study: dynamic and static communication networks matched for total cost (Fig.~\ref{fig1}C).

Once such a temporal network is defined, one can ask how efficiently information can spread across it. We quantified two complementary aspects of communication performance: how many nodes can be reached from a given source within a finite time horizon, and how rapidly destinations are reached on average. Because communication on temporal graphs depends not only on which links exist but also on when they exist, diffusion on these networks differs fundamentally from diffusion on static graphs. In a static graph, an optimal walker follows a shortest path. In a temporal graph, successful transmission instead requires a time-respecting path, that is, a sequence of links that exists in the correct temporal order when each hop is taken \citep{Kempe2002, Holme2012}. Temporal reconfiguration may therefore hinder diffusion, by forcing the walker to wait until an escape link appears, or accelerate it, by creating transient shortcuts unavailable in any single static frame.

To bracket these possibilities, we examined two limiting dispatch strategies (Fig.~\ref{fig1}D). In the \emph{drunk-walk scenario}, packets perform a memoryless random walk and choose uniformly among the currently available outgoing links, with self-loops allowing them to remain in place. In the \emph{smart-walk scenario}, by contrast, walkers have oracle-like access to the future evolution of the temporal network and can therefore select the fastest time-respecting route to each destination. These two limits provide chance-level and upper bounds, respectively, on the communication potential afforded by a given temporal network. Framed this way, the problem becomes determining whether, under equal total cost, redistributing links dynamically across time can improve information dispatch relative to more static allocations of the same communication budget.

This formulation turns thresholded empirical dFC into a controlled toy communication substrate, directly comparable with temporal network null models of increasing complexity and sets up a first question: when links are costly, does temporal reconfiguration hinder information diffusion because links are transient, or improve it because the same budget can be deployed across a more diverse repertoire of communication links?

\subsection{Dynamicity improves information dispatch under costly links}

To answer this question, we compared empirical resting-state temporal networks derived from sliding-window dynamic functional connectivity (dFC) with static counterparts matched for total cost, i.e. for the cumulative number of active links over the observation window. Communication performance was characterized, as introduced above, using two complementary quantitative measures: \emph{irrigation reach}, defined as the number of nodes that can be reached from a source within a pre-fixed observation horizon $T$, and \emph{irrigation resistance}—or penalized latency—defined as a lower-bound summary statistic of the time required to reach target nodes, with unreached targets contributing a penalty of $T+1$. Here we set $T=1200$, corresponding to the number of volumes in long fMRI resting state sessions, typically of this order of magnitude or smaller (see Discussion and Methods). We considered both a smart and a drunk dispatcher, in order to capture both the theoretical optimum and the expectation under random diffusion.

\begin{figure}[H]
\centering
\includegraphics[width=\textwidth]{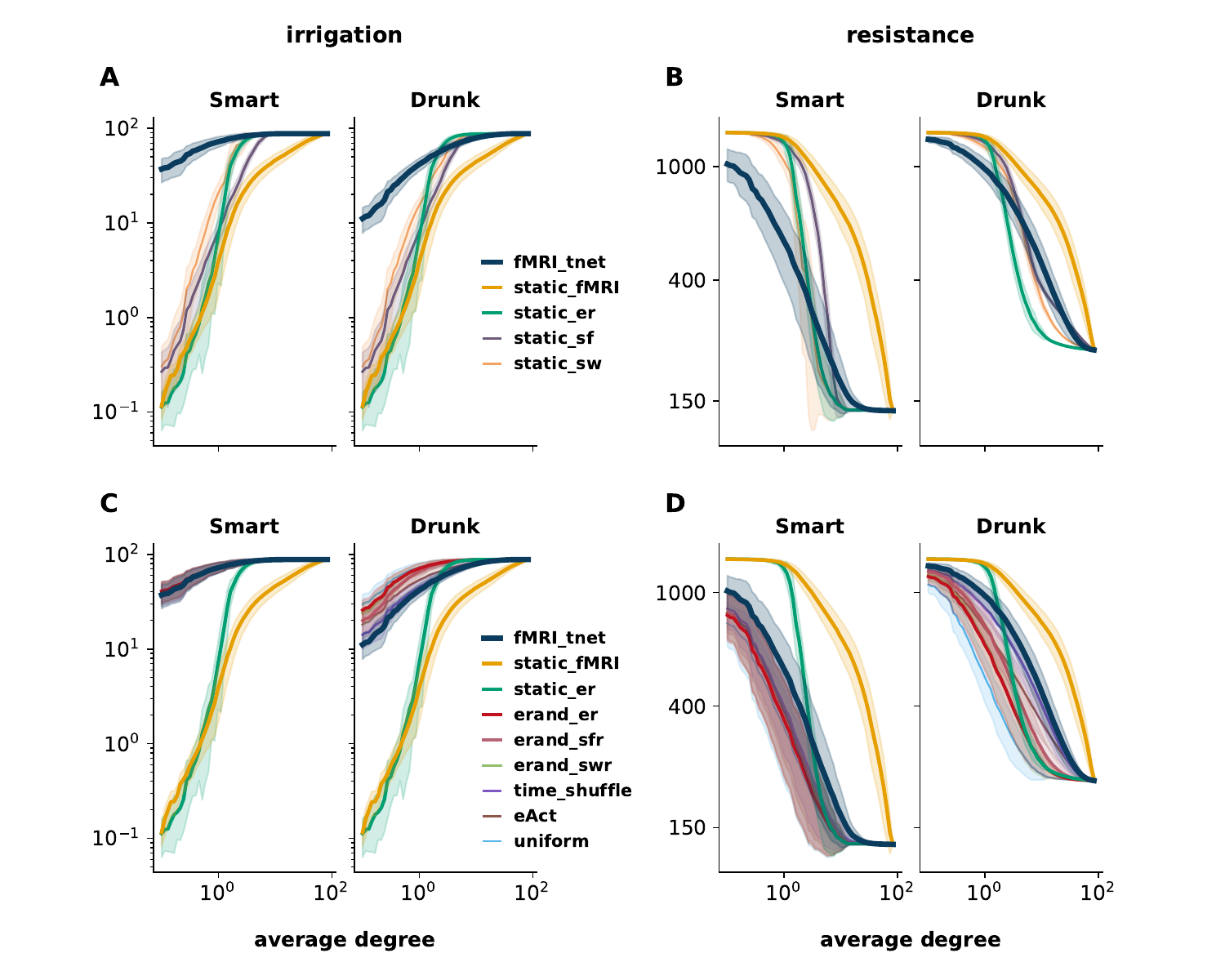}

\caption{\textbf{Dynamic functional connectivity improves information dispatch under costly links.} Communication performance was compared between empirical resting-state temporal networks (\texttt{fMRI tnet}) and equal-cost static controls as a function of average frame degree. Static controls included \texttt{static\_fMRI}, in which all frames were identical to the empirical static FC, and \texttt{static\_er}, \texttt{static\_sf} and \texttt{static\_sw}, in which all frames were fixed quenched random graphs with Erd\H{o}s--Rényi, scale-free and small-world topologies, respectively. A, Irrigation, defined as the mean fraction of the $N$ nodes reachable by an information packet within the observation horizon $T=N$. B, Irrigation resistance, or penalized latency, defined as a lower-bound summary statistic of the time required to reach target nodes, with unreached nodes contributing a penalty of $T+1$. For both metrics, temporal networks outperform equal-cost static controls at low average degree, that is, when maintaining active links is most costly. This advantage progressively vanishes as network density increases and static networks become sufficiently connected. C,D, Comparison of empirical temporal networks with temporal null models preserving different aspects of the spatiotemporal organization of empirical dFC, shown for irrigation (C) and irrigation resistance (D), respectively. These included edge-randomized Erd\H{o}s--Rényi, scale-free and small-world networks with frame-wise degree matched to empirical \texttt{fMRI tnet} (\texttt{erand\_er}, \texttt{erand\_sf} and \texttt{erand\_sw}); a \texttt{time\_shuffle} model preserving empirical frames but randomizing their temporal order; and \texttt{eAct} and \texttt{uniform} models in which links were activated independently at each frame with probabilities set either by empirical linkwise activation rates or by a global mean rate. All temporal null models outperform static controls at equal cost, confirming that temporal reconfiguration alone improves information dispatch under cost constraints. However, several null models also outperform empirical dFC, indicating that resting-state temporal organization does not maximize pure diffusion performance. Left and right subpanels correspond to smart and drunk dispatch, respectively. Curves show mean performance across subjects or realizations; shaded bands indicate 95\% confidence intervals.}\label{fig2}
\end{figure}

Across both dispatch strategies, empirical dFC (\texttt{fMRI\_tnet}) outperformed equal-cost static networks when the average degree of individual frames was low, i.e. precisely in the regime where keeping links active is most costly (Fig.~\ref{fig2}A,B). In this sparse regime, temporal reconfiguration increased irrigation reach and reduced irrigation resistance relative to static Erdős–Rényi (\texttt{static\_er}), static scale-free (\texttt{static\_sf}), and static small-world (\texttt{static\_sw}) controls. This advantage progressively diminished as frame density increased, and all models converged once networks became sufficiently dense that most nodes could be reached even without temporal reconfiguration. Thus, when resources are limited, allocating a fixed budget to transiently reconfiguring links is more effective for information dispatch than investing the same budget in fewer but more persistent connections.

The reduction in irrigation resistance should not, however, be interpreted as meaning that temporal reconfiguration makes successful trajectories intrinsically faster. On the contrary, temporal variability generally introduces a waiting-time effect, because links are only intermittently available and packets may need to pause until a useful edge appears \citep{Holme2012,Starnini2013,Valdano2015}. Supplementary Fig.~\ref{SupFig1} shows this explicitly. When latency is computed only over source-target pairs that successfully communicate within the observation horizon, dynamic temporal networks are not faster than static ones, but instead consistently slower. The gain observed in irrigation resistance therefore arises because temporal reconfiguration strongly reduces the number of pairs that remain disconnected over the observation horizon. In sparse static networks, many such pairs are unreachable and therefore contribute the penalty term $T+1$; in dynamic networks, by contrast, rewiring allows connected components to change membership over time, so that many targets that would remain inaccessible in the static case become reachable at a later frame. Temporal reconfiguration therefore improves irrigation resistance not by accelerating local diffusion along realized paths, but by reducing communication failure enough to outweigh the delay induced by intermittency (i.e. they reduce \emph{inaccesibility} more than they increase \emph{---unpenalized--- latency}, see Supplementary Fig.~\ref{SupFig1}).

The same mechanism also explains the increase in irrigation reach. In sparse static networks, communication is limited by the connected component containing the source. Below or near percolation—$\langle k\rangle \sim 1$ for Erdős–Rényi graphs \citep{ER1960}—this component excludes a substantial fraction of the network, so that even an optimal dispatcher cannot reach all targets. Temporal reconfiguration relaxes this constraint by making the accessible component evolve over time: as links change, different nodes enter and leave the region that can be reached from the source through time-respecting paths. A fixed budget of costly links can therefore sample a much larger fraction of the network across time than any static allocation with the same total cost. This effect is strongest in the smart case, where dispatch can exploit the exact timing of transient shortcuts and bridges, but remains substantial in the drunk case, showing that the communication gain does not require explicit optimization of routes.

We next asked whether the gain observed in empirical dFC merely reflects the generic benefit of temporal variability, or whether the specific spatiotemporal organization of resting-state dFC carries additional consequences for communication. To address this question, we compared empirical temporal networks with a family of null models preserving different subsets of their spatial and temporal statistics (Fig.~\ref{fig2}C,D). More specifically, we constructed a hierarchy of null models that selectively preserved particular spatial and temporal features of the empirical temporal networks while randomizing others (see Methods for details). These nulls ranged from minimally constrained ensembles (\texttt{uniform}), in which links were redistributed with little reference to the empirical spatiotemporal organization, to more structured surrogates preserving framewise densities (\texttt{erand\_er}, \texttt{erand\_sf}, \texttt{erand\_sw}), linkwise activation rates (\texttt{eAct}, exact network frames but with disrupted temporal sequence \texttt{time\_shuffle}), or combinations of link rewiring and frame shuffling (\texttt{erand\_and\_time}). This allowed us to ask whether communication performance depends primarily on the overall amount of temporal variability, on the heterogeneity of link activation probabilities, on the preservation of within-frame spatial structure, or on the sequential ordering of frames.

All temporal null models outperformed static controls, confirming that temporal reconfiguration alone is sufficient to improve information dispatch at equal cost. However, many null models also outperformed empirical dFC itself, in some cases markedly so, especially at low average degree. This held for both irrigation reach and irrigation resistance, and under both smart and drunk dispatch, although the relative ordering of null models depended somewhat on the metric and dispatch strategy considered. In general, the more random and less structured the temporal reconfiguration was, the greater the irrigation reach and the lower the irrigation resistance. Consistently, in purely random temporal networks where each frame was generated from a common reference graph (Erdős–Rényi, scale-free or small-world) by applying a prescribed fraction of random, uncorrelated rewiring, irrigation reach increased and irrigation resistance decreased monotonically with the rewiring fraction (Supplementary Fig.~\ref{SupFig2}). Thus, while empirical resting-state dFC clearly benefits from dynamicity, it does not maximize pure diffusion performance.

Taken together, these results show that empirical dFC supports communication better than any equal-cost static architecture, yet remains less efficient than it could be if maximizing diffusion were the sole objective. This suggests that resting-state dynamicity is shaped not simply to maximize global spreading, but to satisfy a broader organizational compromise, which we next characterize in terms of the joint maintenance of integration, through long-range information propagation, and segregation, through the transient confinement of information within restricted local neighborhoods.

\subsection{Empirical dynamic functional connectivity transiently preserves local segregation}

Efficient brain function requires not only the integration of information across distributed regions, but also the transient preservation of local processing within partially segregated subnetworks \citep{Tononi1994, Sporns2013}. In the previous section, we showed that empirical dynamic functional connectivity (dFC) improves integrative communication under costly links by enhancing irrigation (increased reach and reduced resistance relative to equal-cost static networks). We next asked whether this gain in integrative spreading comes at the expense of segregation, or whether empirical resting-state dFC also preserves spatiotemporal features that transiently confine information within local neighborhoods before broader dispersal.

\begin{figure}[H]
\centering
\includegraphics[width=\textwidth]{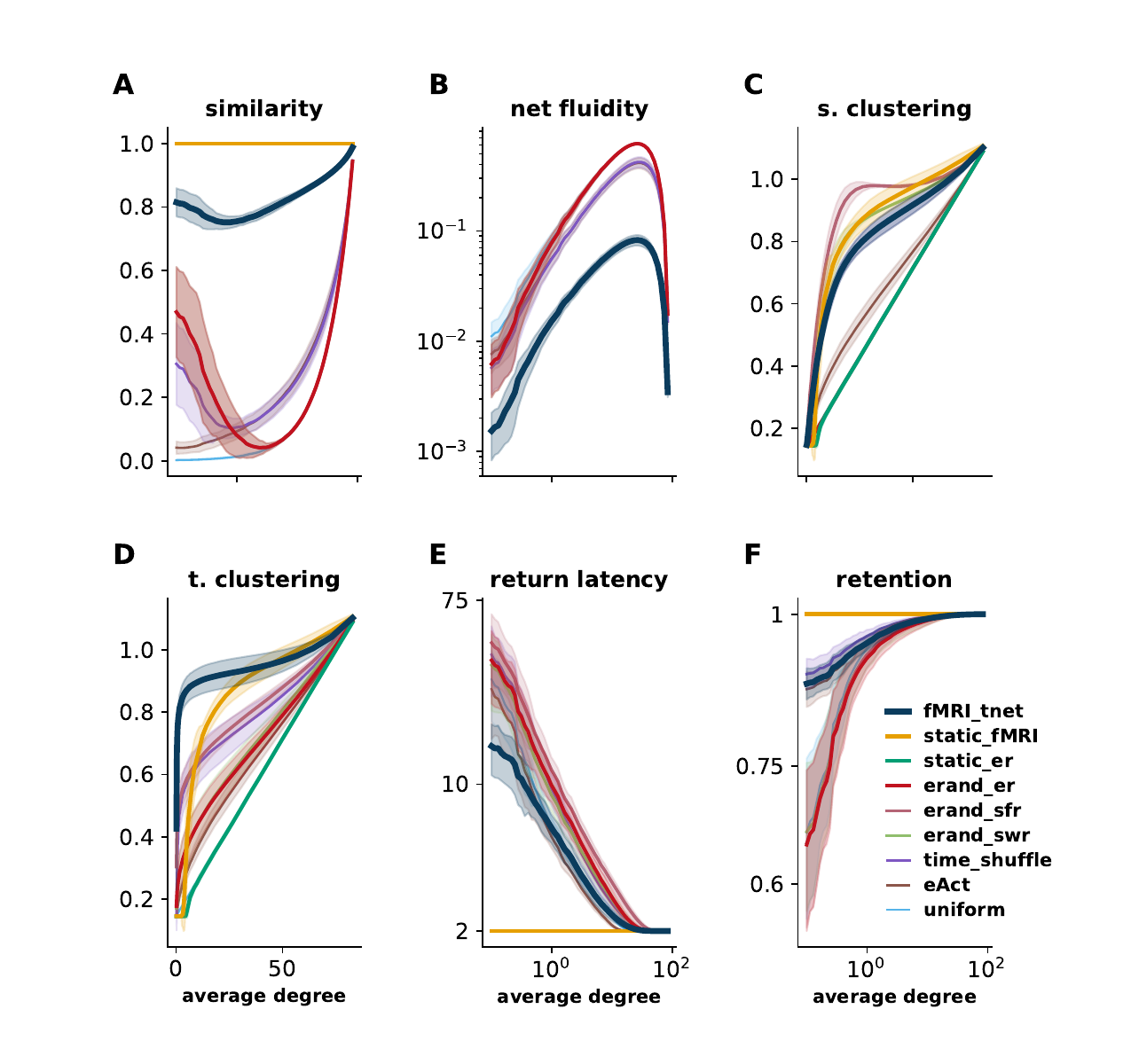}
\caption{\textbf{Empirical dynamic functional connectivity promotes transient segregation of information spreading.} Empirical temporal networks were compared with static and temporal null ensembles matched in cost but preserving different subsets of the empirical spatiotemporal structure. Model labels are as in Fig.~\ref{fig2}. A, Cosine similarity between consecutive frames, showing that empirical dFC evolves more smoothly over time than any temporal null model and remains closest to the static limit. B, Net fluidity of link changes across consecutive frames, defined as the product of frame-to-frame and overall variability measures, showing that empirical dFC follows a more constrained and predictable temporal evolution than more randomized nulls. C, Mean spatial clustering of individual frames. In the sparse regime, empirical dFC frames are more clustered than static controls, which remain closer to tree-like configurations below percolation. D, Temporal clustering coefficient between consecutive frames ($\Delta = 1$), showing that empirical dFC more strongly preserves local cohesive structure across time than any other null model. E, Return latency, defined as the mean shortest time required for a packet that has left a source node to return to it under drunk dispatch. F, Retention, defined as the fraction of neighbors present at time $t+1$ that were already neighbors at time $t$. Empirical dFC exhibits shorter return latency and higher retention than more randomized temporal nulls, again approaching the static limit. Together, these results show that empirical dFC does not simply favor global spreading, but also transiently stabilizes locally cohesive neighborhoods, thereby supporting segregation before later large-scale integration.}\label{fig3}
\end{figure}

A first indication comes from the rate and regularity of network reconfiguration. We quantified ``statism'', i.e. the similarity between consecutive frames by the \emph{cosine similarity} between their binarized adjacency matrices, viewed as edge vectors. Large cosine similarity means that the active-link pattern at time $t+1$ remains close to that at time $t$, i.e. that network reconfiguration is relatively slow. Empirical dFC exhibited substantially higher cosine similarity than any temporal null model considered here, remaining closest to the static limit across costs (Fig.~\ref{fig3}A). As a complementary compact summary descriptor of ``dynamism'', we also considered \emph{Net fluidity}, a composite index obtained by multiplying global edge-activity entropy, measuring global irregularity of network reconfiguration, by average dFC speed, i.e. the average rate of frame-to-frame variability, as defined by \citet{Battaglia2020}. Net fluidity analyses in Fig.~\ref{fig3}B again place empirical dFC in a less globally unstable regime than more randomized temporal nulls.

This slower reconfiguration is accompanied by stronger local cohesiveness in space and time. At the level of individual frames, empirical dFC displayed stronger \emph{spatial clustering} than static controls in the sparse, high-cost regime (Fig.~\ref{fig3}C). Here spatial clustering refers to the usual clustering coefficient computed within each frame, and therefore measures the abundance of closed local neighborhoods at a given instant. This is notable because equal-cost static networks below or near percolation tend to organize into relatively tree-like components, poor in closed local motifs. By contrast, empirical dFC frames retain richer local triangular structure. Time-shuffled controls, as expected, preserve the same framewise spatial clustering, since shuffling frame order leaves the topology of each frame unchanged.

We then asked whether this local cohesiveness also persists across time. Because standard clustering is defined on a single graph, we introduced a \emph{temporal clustering} measure based on clustering within short $\Delta$-aggregate network. For $\Delta=1$, this means aggregating two consecutive frames, $t$ and $t+1$, and measuring how much closed local structure is present in this minimal space-time neighborhood. Temporal clustering is therefore high when local motifs are not necessarily complete within every time frame but appear when combining multiple consecutive frames. This temporal clustering notion is related to previous notions of temporal clustering across adjacent frames and to multilayer generalizations of triangle-based transitivity, but is tailored here to capture the persistence of quasi-clustered neighborhoods over short time scales \citep{Tang2009, Cardillo2014, Cozzo2015, Bail2024}. Empirical dFC displayed higher temporal clustering than all null models considered here (Fig.~\ref{fig3}D), indicating that local cohesive structure is propagated more strongly across time than in more randomized surrogates. In the watery metaphor of our information-dispatch game, this corresponds to a local ``slush'': the flow is not immediately flushed away into the whole system, but remains transiently caught within overlapping local semi-frozen pockets.

These structural features have direct consequences for the short-timescale confinement of information. We first quantified \emph{return latency}, defined as the shortest theoretical time needed for a packet that has left a source node to be able to return to it, averaged over time frames and nodes. Importantly, this is not a random-walk quantity, but a semi-analytically computed lower-bound derived from successive temporal-network products (see \emph{Methods}). It captures how rapidly local recirculation channels can in principle re-open around a source, averaged across nodes and starting times. Empirical dFC showed shorter return latency than most temporal null models, again approaching the static limit (Fig.~\ref{fig3}E). We also quantified \emph{retention}, defined as the fraction of the neighborhood present at time $t+1$ that was already present at time $t$. Large retention indicates that local neighborhoods renew slowly and preserve substantial overlap from one frame to the next. Empirical dFC showed consistently higher retention than more randomized temporal nulls (Fig.~\ref{fig3}F). 

The null-model comparison helps clarify which ingredients matter most. The least constrained temporal ensembles, such as \texttt{uniform}, in which frames are generated independently with spatially uniform and temporally constant link probability, are also the least segregative: they show the lowest frame-to-frame similarity, the weakest clustering, the longest return latency, and the lowest retention. Ensembles such as \texttt{erand\_er} and related scale-free \texttt{erand\_sfr} or small-world \texttt{erand\_swr} random-frame families are likewise less segregative and less clustered than the empirical data, although their detailed ranking depends on the metric (cf. summary table in Fig.~\ref{fig5}). Thus, increasing temporal randomness may improve integrative spreading, as shown in the previous section, but it weakens the transient preservation of local neighborhoods.

Taken together with the previous section, these results show that empirical dFC supports a genuine integration–segregation compromise. Relative to equal-cost static architectures, it improves long-range irrigation; relative to more randomized temporal surrogates, it preserves stronger local persistence, stronger space-time clustering, shorter return latency, and higher neighborhood retention. Empirical dynamicity therefore does not merely allow information to spread far: it also allows it to linger, recirculate, and transiently eddy within local neighborhoods before being dispersed network-wide.

These non-trivial segregation properties call for an investigation of the mechanisms that may lead to them. If empirical resting-state dFC preserves neighborhoods more than generic temporal nulls, this property must arise from specific organizing principles of its spatiotemporal structure rather than from unstructured dynamicity alone.

\subsection{Staticized null models and connectome-based dynamics partly explain neighborhood-preserving dFC}

The previous sections showed that empirical resting-state dynamic functional connectivity (dFC) occupies an intermediate regime between equal-cost static architectures, which are less integrative and more segregative, and more generic temporal null architectures, which are more integrative and less segregative. This intermediate regime appears to depend on the combination of dynamicity with non-trivial neighborhood preservation. We next asked where this compromise comes from.

\begin{figure}[H]
\centering
\includegraphics[width=\textwidth]{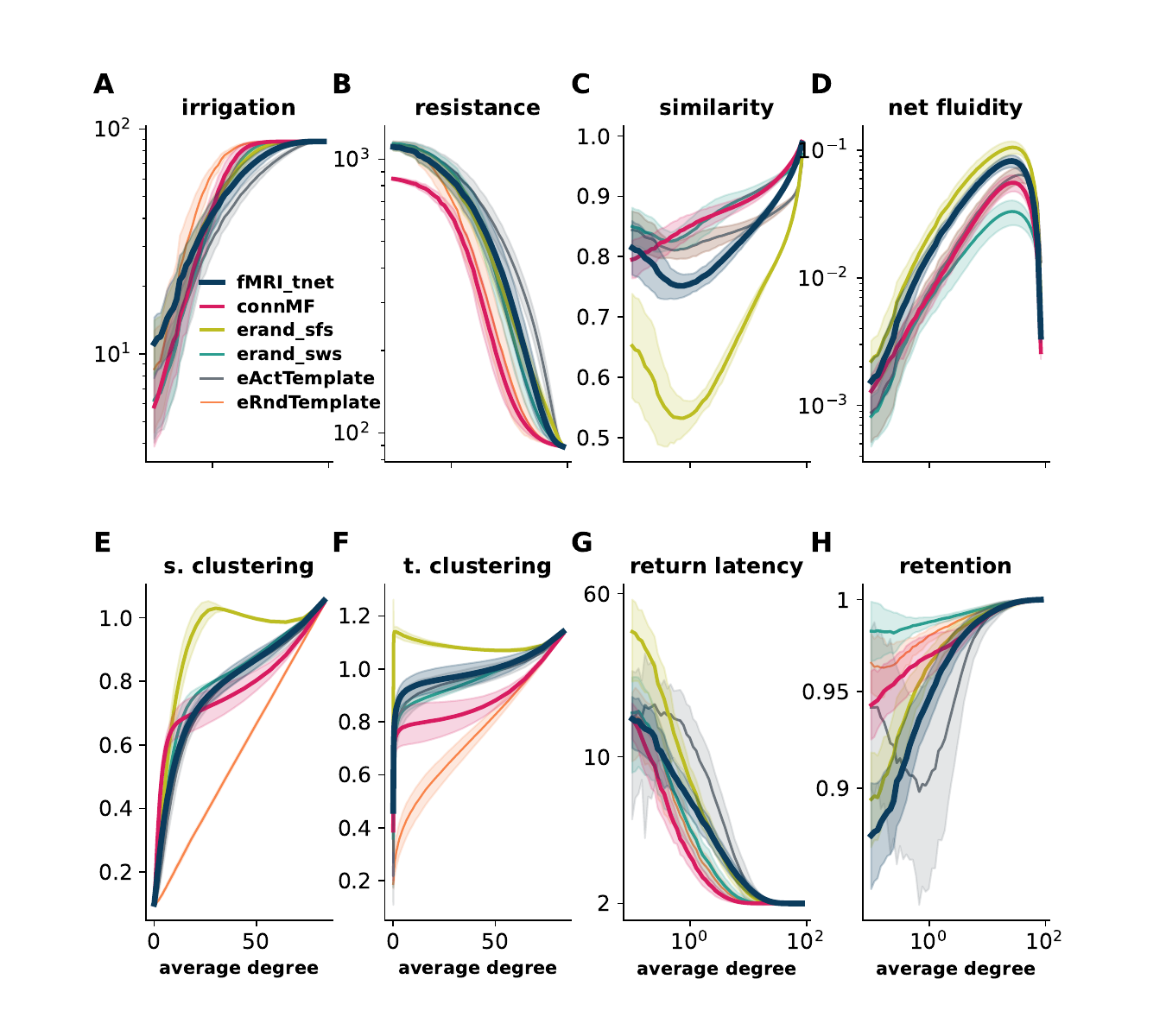}
\caption{\textbf{Staticized null models and connectome-based dynamics partly explain the quasi-staticity of empirical dFC}. Empirical temporal networks were compared with ad hoc null models in which temporal persistence is explicitly reinforced, as well as with dFC streams generated by simulated resting-state activity in a connectome-based mean-field model (\texttt{ConnMF}). The staticized nulls include \texttt{eActTemplate}, in which links are activated according to a fixed priority ranking derived from empirical activation probabilities; \texttt{randTemplate}, in which the same priority-template mechanism is used with a randomly chosen quenched ranking; \texttt{rand\_sfs}, in which scale-free hub identity is frozen across frames; and \texttt{rand\_sws}, in which neighborhood lattice structure is frozen across frames. A, Irrigation reach. B, Penalized latency. C, Cosine similarity between consecutive frames. D, Net fluidity of link changes. E, Spatial clustering. F, Temporal clustering. G, Return latency. H, Retention (cf. Figs.~2 and 3). Across panels, all staticized models move closer to the empirical regime than previously considered random temporal nulls, confirming that reinforced spatiotemporal correlations recover part of the quasi-staticity of empirical dFC. However, none of the hand-built template models fully reproduces the empirical combination of integration, transient segregation, and local clustering. By contrast, \texttt{ConnMF} captures, at least qualitatively, several nontrivial features simultaneously, including high spatial and temporal clustering together with strong and simultaneous integrative and segregative performance. It nevertheless remains more persistent than the empirical data in terms of frame-to-frame similarity.}\label{fig4}
\end{figure}

To address this question, we first constructed a family of ``staticized'' null models in which temporal variability no longer arises solely from independent stochastic edge activations, but instead unfolds around a persistent scaffold. The aim was to test whether empirical dFC could be explained by a partially frozen template, that is, by a situation in which the activation of some links or nodes is consistently prioritized across frames and temporal variability arises only from fluctuations around this prioritized substrate.

The first such model, \texttt{eActTemplate}, extends the previous \texttt{eAct} null by introducing a fixed priority order among edges. Links are ranked according to their empirical activation probabilities, and each frame is generated by activating edges in that fixed order until the required number of links for that frame is reached. Dynamicity therefore still arises from framewise degree fluctuations, but the order of link recruitment remains constant across time. Highly ranked edges are repeatedly selected and thus become much more persistent than lower-ranked ones, yielding a quasi-static backbone of preferred interactions. We also considered a control version of this construction, \texttt{randTemplate}, in which the priority order is chosen randomly once and then kept fixed across all frames. This model imposes the same type of quenched persistence, but without aligning the persistent backbone with empirically preferred links.

We extended the same logic to random graph ensembles with non-trivial topology. In \texttt{erand\_sfs}, the scale-free degree hierarchy is frozen, so that hub identity remains stable across frames. In \texttt{erand\_sws}, neighborhood ordering is frozen, so that local lattice relations are partially preserved through time. These models therefore introduce stronger temporal correlations than the previously considered \texttt{erand\_sfr} and \texttt{erand\_swr} nulls, in which scale-free or small-world structure is regenerated more freely from frame to frame.

All of these staticized null models moved substantially closer to the empirical data than the earlier random temporal surrogates (Fig.~\ref{fig4}). Relative to the previous nulls, they displayed higher cosine similarity, lower global dynamism, stronger spatial and temporal clustering, shorter return latency, and higher retention, all consistent with greater persistence of local interaction structure across time. This shows that reinforced temporal correlations are sufficient to recover an important fraction of the quasi-staticity that characterizes empirical dFC.

However, none of these constructions fully reproduced the empirical pattern. The \texttt{eActTemplate} model captured part of the neighborhood-preserving organization, including stronger spatial and temporal clustering than more weakly constrained nulls, but remained both less integrative and less segregative than the empirical temporal networks: irrigation reach remained lower, irrigation resistance higher, and local recirculation and retention weaker than in the data. Thus, enforcing persistence through an empirically ranked edge hierarchy enhances local cohesiveness, but at the cost of overconstraining the system and limiting both global spreading and local return. The \texttt{randTemplate} model behaved differently. It could achieve strong communication performance together with relatively strong return and retention, but did so without reproducing the structured local organization of empirical dFC. Because its persistent backbone is spatially random, it did not recover the spatial or temporal clustering profile of the data. It therefore reproduced some functional consequences of persistence without accounting for the topology through which they arise.

The same conclusion applies to the staticized scale-free and small-world surrogates. Freezing hub identity in \texttt{rand\_sfs} or neighborhood order in \texttt{rand\_sws} increased temporal persistence relative to their fully re-randomized counterparts, but neither model simultaneously matched the empirical levels of integration, segregation, and local clustering. These results indicate that partial freezing of graph topology can reproduce some aspects of empirical quasi-staticity, but that quasi-staticity alone is not sufficient. What matters is not only that some structure persists, but which structure persists and how that persistence is embedded in an evolving temporal network.

We then asked whether a less artificial mechanism, grounded in neural population dynamics constrained by a realistic anatomical substrate, could better account for the observations. For this purpose, we turned to mechanistic simulations of resting-state activity based on connectome-based mean-field dynamics \citep{Deco2013}, and extracted a virtual dFC stream from simulated resting-state BOLD signals matched in duration to the empirical data. We considered specifically a Mean-Field model by \citet{Pathak2025} with temporally fluctuating working-point parameters (here labeled \texttt{ConnMF}), previously shown to reproduce broad descriptors of empirical dFC variability and to outperform earlier state-of-the-art models \citep{Hansen2015}. Strikingly, \texttt{ConnMF} recovered several non-trivial features of the empirical integration–segregation compromise. It reproduced the high spatial clustering of empirical dFC, plausibly reflecting in part the clustered organization of the structural connectome, but also the high temporal clustering, which is more informative because no explicit edge-priority rule had been imposed. Unlike the template-based nulls, \texttt{ConnMF} generated space-time cohesive dynamics through self-organization alone, constrained by the underlying anatomy, without being forced to preserve any particular ordering of links.

At the same time, \texttt{ConnMF} still differed systematically from the empirical data. It was more integrative, with higher irrigation reach and lower penalized latency, but also more segregative, displaying shorter return latency and higher retention than the empirical temporal networks. By the present metrics, \texttt{ConnMF} therefore realized a joint boosting of integration and segregation even stronger than in the empirical data. This distinguishes it from the artificial null models: \texttt{eActTemplate} improved clustering but remained both less integrative and less segregative than the data, whereas \texttt{randTemplate} improved communication and return, but in a largely unstructured manner. \texttt{ConnMF}, by contrast, is the only model in this set that combines reproduction of substantial spatial and temporal clustering with strong integration and strong transient segregation.

The remaining discrepancy is nonetheless informative. In particular, empirical dFC shows lower frame-to-frame similarity than would be expected from the neighborhood-preserving null models and from \texttt{ConnMF} as well. In other words, the data preserve local structure without becoming excessively static. This suggests that empirical dFC may involve an additional decorrelating mechanism that renews frame topology sufficiently from one time point to the next (see Discussion), while still preserving enough temporal organization in the fluctuations to sustain clustering, return, and retention.

\begin{figure}[H]
\centering
\includegraphics[width=\textwidth]{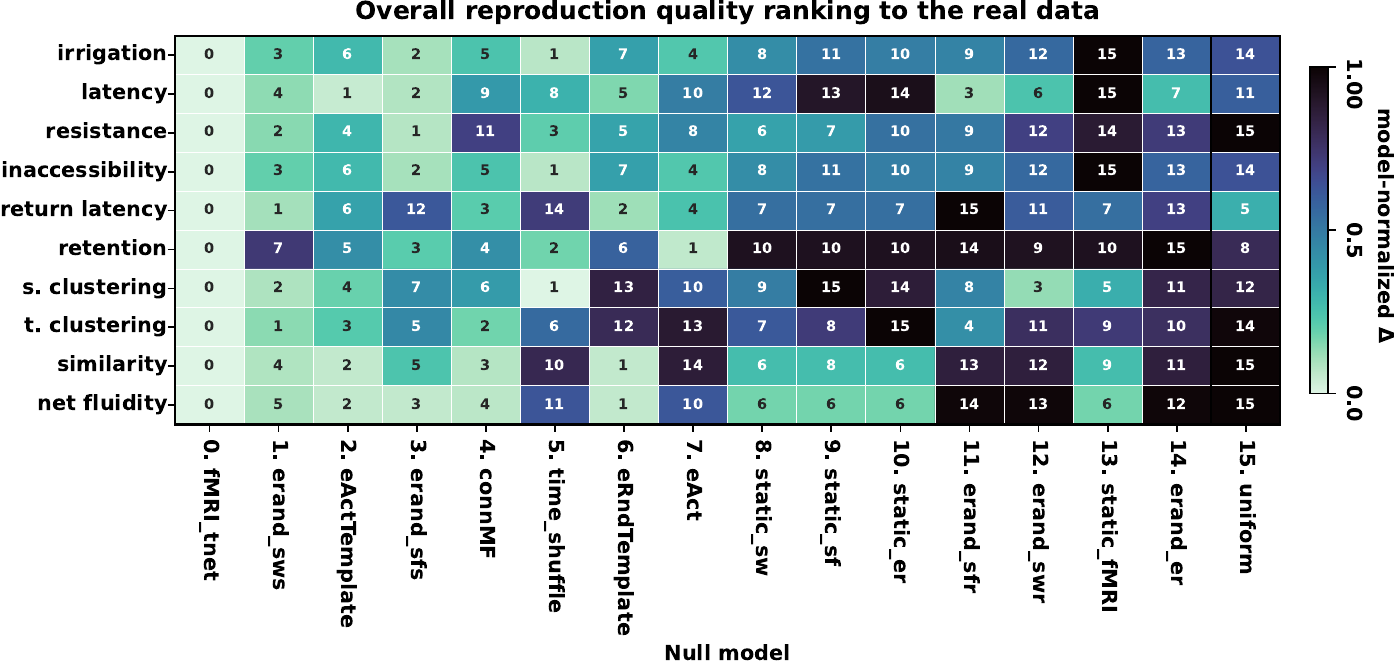}
\caption{\textbf{Overall ranking of null models by similarity to empirical temporal functional connectivity.} Heat map summarizing the agreement between empirical resting-state temporal networks and the different null models across the full set of communication, segregation and temporal-organization metrics. Rows correspond to the different features considered, and columns to null models (all model labels are as in Figs.~\ref{fig2} and~\ref{fig4}). For each feature and model, agreement was quantified by a model-normalized $\Delta$, measuring the discrepancy between the link-cost-dependent curve of the null model and the corresponding empirical \texttt{fMRI tnet} curve; larger normalized $\Delta$ values indicate greater deviation from the empirical data. Cell color encodes the normalized $\Delta$, whereas the number displayed in each cell gives the rank of the corresponding model for that feature, from best match (0) to worst match (15). Columns are ordered by median rank across all features, yielding an overall performance ranking. The closest matches to empirical dFC are provided by the staticized models \texttt{erand\_sws}, \texttt{eActTemplate} and \texttt{erand\_sfs}. \texttt{ConnMF} also ranks among the top-performing models and is notable as the only physiologically interpretable model that simultaneously reproduces, with reasonable accuracy, integration, segregation and temporal-clustering properties, while generating these features through self-organized dynamics rather than an artificial explicit construction.}\label{fig5}
\end{figure}

Taken together (see Fig.~\ref{fig5} for a final ranking of ``top'' and ``poorer'' performer models and Fig. \ref{SupFig3} for a summary synoptic view of all metrics), these results substantially narrow the space of possible explanations. Empirical dFC cannot be accounted for by generic temporal randomness, but neither is it captured by a simple partially frozen template. Instead, it appears to require a structured dynamical regime in which local neighborhoods persist sufficiently to support transient recirculation and clustering, yet are actively decorrelated enough to avoid excessive statism. Among the models considered here, connectome-based mean-field dynamics come closest, at least qualitatively, to this regime, suggesting that the coexistence of integration and transient segregation may emerge naturally from coordinated collective dynamics unfolding on a realistic anatomical scaffold, but that additional mechanisms are still required to reproduce the exact empirical balance.

\section{Discussion}

Struck by the ubiquity of dynamic reconfiguration in brain functional connectivity across spatial and temporal scales, we asked whether such dynamicity might serve a functional purpose rather than simply reflecting noise or an incidental by-product of ongoing activity. Using information spreading on empirical temporal networks of resting-state fMRI functional connectivity as a representative case study, we found that the “d” in dFC is not generically advantageous. Yet we also found that it can become strongly beneficial under specific operating conditions: when establishing, maintaining, and using communication links is costly, such that inter-regional exchange must proceed through very sparse instantaneous functional networks. In this regime, revealed by strong thresholding and marked dilution, empirical dFC is clearly not equivalent to random rewiring, measurement noise, or trivial sampling variability. Rather, its temporal organization supports propagation and integration better than staticized or time-scrambled controls, while preserving nontrivial aspects of segregation. Thus, the pervasive dynamicity of FC need not be viewed simply as a nuisance to be averaged away. It may instead reflect an organized regime in which temporal reconfiguration becomes useful precisely because a single static frame is too poor, too fragmented, or too costly to sustain efficient communication on its own. Although this does not prove that the brain optimizes dFC for communication, it does show that, if communication had to operate under stringent budget constraints, structured dynamicity may become a genuine functional asset.

This point is particularly interesting when placed against the background of temporal-network statistical physics. A large statistical-physics literature has shown that temporality often hinders diffusion relative to static aggregates, because the ordering of contacts breaks time-respecting paths, lowers effective transitivity, and slows exploration compared with the corresponding time-collapsed network \citep{Starnini2013, Masuda2013}. Yet that same literature has also shown that temporal ordering is not generically detrimental: memory and causal correlations can either slow down or accelerate diffusion \citep{Scholtes2014, Delvenne2015}, and temporality can yield important advantages for control and routing \citep{Li2017}. What distinguishes the present case is the operating regime. We explicitly introduce a cost principle and examine spreading in ultradiluted functional networks, where strong thresholding renders individual frames sparse enough that many communication opportunities are simply absent at any given instant. In such a regime, temporal reconfiguration is no longer merely a source of path disruption relative to an already well-connected static substrate. It can instead become the very mechanism by which disconnected opportunities are sequentially stitched into viable routes. The advantage we observe therefore does not stand in opposition to the temporal-networks literature; rather, it reveals a complementary regime in which the effect of dynamicity can flip sign, because communication is evaluated not against a dense aggregate but under stringent budget constraints and near-fragmented instantaneous connectivity.

In such a high-cost ---or equivalently, low-budget--- regime, sharp thresholding isolates only the strongest instantaneous functional interactions, and these are unlikely to constitute a random residue of the full FC matrix. As shown by \citet{Pajevic2012}, many real weighted networks, including brain-related networks, display an integrative weight organization whereby strong links preferentially connect nodes with highly overlapping neighborhoods. In such networks, clustering remains strikingly robust to the removal of weak links, but collapses rapidly when strong links are pruned. This provides a natural framework for our findings. It may account not only for the elevated clustering observed in empirical dFC relative to more random null models, but also for its temporal clustering. Indeed, fluctuations of link weights above or below a sharp threshold may progressively reveal different portions of an underlying clustered strong-link scaffold, much like the successive emergence of the tip of an iceberg whose cohesive substructure remains mostly hidden below threshold at any given time. At the same time, weak links should not be regarded as irrelevant leftovers. \citet{Pajevic2012} argued that strong interactions support robust communication, whereas weak interactions contribute a more exploratory organization. In our setting, transient excursions of weaker links above threshold may therefore broaden integrative reach and lower communication resistance. This suggests that FC may be better understood not simply as a dense weighted network, but as the superposition of two interacting infrastructures: strong links forming transient highways, embedded within a weaker and more diffusive sea. In the brain, this weaker background may remain functionally important, because polysynaptic communication and communicability over many indirect and individually weak routes can shape functional organization beyond direct pathways alone \citep{Goni2014,Seguin2022}. Its dynamic interplay with the stronger scaffold ---eventually informed by higher-order correlations \citep{Faskowitz2019, Lombardo2020, Arbabyazd2023}--- may further modulate, stabilize, or destabilize over time which links effectively cross the threshold for costly activation.

In this study, we treated FC as a substrate for information diffusion. At first sight, this choice may seem counterintuitive, because FC has no direct material existence: rather than describing physical connections, it captures coordination among activity signals. For this reason, most communication models in brain network science have been formulated on the structural connectome, which naturally defines the substrate of possible signal transmission \citep{Avena2018,Seguin2023} and also constrains the propagation of externally injected stimulation \citep{Seguin2023stim}. However, anatomical possibility does not necessarily translate into effective transfer. A structural edge may exist without supporting communication in the current dynamical state \citep{Battaglia2012,Kirst2016}, whereas FC captures which interactions are effectively coordinated and statistically open at a given moment, including interactions mediated by polysynaptic pathways that may act as effective zero-lag shortcuts despite the absence of direct anatomical links \citep{Vicente2008}. In correlation-based FC, this interpretation is not merely heuristic: correlation reflects the linear component of mutual information—often a dominant contribution \citep{Hlinka2011}—and thus provides a conservative proxy for communication potential in an information-theoretical sense \citep{ShannonWeaver1949}. FC can therefore be viewed as defining the communication geometry dynamically available on top of the structural substrate, compactly integrating the joint contributions of mono- and polysynaptic pathways together with their modulation by ongoing dynamics. Crucially, such a geometry can shape system-level behaviour: when dynamics are dominated by collective coordination, the distributed nonlinear effects of focal stimulation are better predicted by shortest-path distance on the functional connectome than on the structural connectome \citep{Papadopoulos2020}. Treating FC as an emergent, dynamic infrastructure for information dispatch is therefore not only plausible, but mechanistically justified.

Work on resting-state activity has long shown that large-scale FC reflects transitions across a structured repertoire of states \citep{Hutchison2013,Calhoun2014,Preti2017}—likely corresponding to the sampling of the system’s “dynome” \citep{Deco2013,Hansen2015}—rather than the expression of permanently active networks. What our results add is that, under strong thresholding and explicit budget constraints, this dynamical view implies not merely variability, but genuine intermittence: only a small fraction of links remain effectively open at any given moment, and substantial periods of disconnection become intrinsic to the communication architecture. Although this may appear radical, it is not without precedent. In the classical communication-through-coherence framework, phase synchrony opens communication channels by aligning windows of excitability, but the implicit picture often remained one of relatively sustained coherence supporting information exchange \citep{Fries2005,Fries2015}. More recent work has shifted this view toward a markedly more intermittent regime, in which flexible routing is achieved through brief, self-organized bursts of synchrony unfolding despite stochasticity, rather than through continuously open channels \citep{Palmigiano2017}. Our results are fully consistent with that reorientation. If maintaining a functional communication channel is costly, then communication need only become effectively open when locally processed information is ready to be flushed into larger-scale interactions. 

Beyond the reduced control burden associated with steering small transient network motifs rather than persistently open channels \citep{Li2017}, bursty coupling may also be energetically advantageous. In that sense, the link-cost parameter in our model need not be regarded as a purely abstract graph-theoretical penalty, but may admit a concrete physiological interpretation. Establishing and maintaining functional coupling may carry substantial biological costs, including the energetic demands of spiking, synaptic transmission, and restoration of ionic gradients \citep{Attwell2001,Howarth2012,Hallermann2012}, as well as the additional expenditure required to sustain coordinated population activity—such as the oscillatory background on which messages to broadcast, for example in the form of finer rate modulations, may ride \citep{Buzsaki2007,Kann2012}. More broadly, intrinsic functional coordination appears to consume a substantial “dark energy” fraction of the brain’s metabolic budget \citep{Raichle2006,Zhang2010}, and PET studies suggest that FC itself carries measurable energetic costs, particularly in hub regions that sustain many interactions over time \citep{Tomasi2013,Volpi2024}. From this perspective, the sparse and temporally structured regime revealed here may be not a mathematical curiosity, but a plausible physiological strategy for sustaining rich information dispatch at limited energetic cost.

The extensive comparison with a hierarchy of null models of increasing structure makes clear that the temporal organization of empirical dFC cannot be reduced to a merely bursty, stochastic-like activation of links occurring independently of one another (cf. the marked mismatch between \texttt{fMRI tnet} and the \texttt{eAct} null in Figs.~\ref{fig2}~and~\ref{fig3}). Among its least trivial features—and the most difficult for null models to reproduce jointly with strong integrative and segregative performance—is the elevated temporal clustering observed especially in the high-cost regime, pointing to a rich underlying dynamical process. Across the different notions of temporal clustering proposed in the literature ---ranging from persistence of a node’s neighborhood over time \citep{Nicosia2013} to explicit tracking of temporally closed triangles \citep{Cui2013}--- our original measure instead relies on ``$\Delta$-aggregation'', i.e. the merging of links active across a short window of $\Delta$ consecutive frames \citep{Holme2012,Krings2012}. Clustering in a $\Delta$-aggregate does not imply that individual frames are themselves clustered. Rather, it reveals the loose maintenance of cohesive mesoscale ensembles: sets of nodes that remain, for some time, more mutually accessible and more jointly organized than chance would predict, even though their precise constituent links fluctuate from frame to frame. In that sense, our measure is closer in spirit to approaches probing temporally metastable and densely but loosely wired structures such as temporal cores or temporal rich clubs \citep{Ciaperoni2020,Pedreschi2022}.

A plausible generator of such transiently cohesive ensembles is active propagation. Rather than reflecting only passive diffusion, temporal clustering may index thresholded or cascade-like recruitment, whereby incoming activity pushes recipient nodes above threshold and initiates further propagation \citep{Nematzadeh2014}. Repeated patterns of simultaneous or sequential brain activation \citep{Mitra2015,Liu2018}, together with critical avalanche-like dynamics triggered by discrete activity fluctuations \citep{Tagliazucchi2012} and global signal pulsation \citep{Glomb2018}, provide possible candidate mechanisms for the emergence of temporal clustering of active links in empirical dFC. From this perspective, the strong temporal clustering of high-cost dFC links may reflect the transient waxing, propagation, and waning of coordinated activity packets. This suggests a concrete direction for future work: testing whether the distributions of cluster size, lifetime, and growth or decay profile exhibit scaling laws consistent with operation near criticality \citep{Hengen2025}, together with its associated computational benefits \citep{Shew2011}. More specifically, operating near the subcritical edge of a percolation transition could naturally generate connected graph fragments spanning a broad range of sizes \citep{Newman2007,Cirigliano2024}, which may then be temporally concatenated into viable communication routes without requiring the sustained maintenance of a costly giant component.

Approaching the conclusion, several limitations of our work should be made explicit. Our analysis assumes that network reconfiguration and spreading occur on matched timescales, a useful modeling choice but not a demonstrated biological fact. If propagation is slower, the framework approximates diffusion over short temporal aggregates; if faster, individual frames become more important, potentially increasing departures from shuffled controls while reducing the net gain from dynamicity. The matched-timescale case should therefore be regarded as a canonical point within a broader multiscale theory, not as the uniquely correct physiological regime. This is a reasonable starting point, given that brain activity is structured across many timescales, from scale-free EEG microstate sequences \citep{VandeVille2010} to cortical reservoirs of intrinsic time constants stretching over multiple decades of durations \citep{Bernacchia2011}. In addition, our resistance measure depends on the observation horizon $T$: choosing $T$ equal to the number of volume frames in typical empirical fMRI sessions offers a pragmatic compromise, but other choices would alter the balance between latency and reachability. Thus, the present results identify a robust computational principle, while falling short of a full reconstruction of message propagation in the brain.

One of the clearest lessons of this work is that, under high link cost or low communication budget, integration is easier to obtain than segregation. Global reach can be enhanced by relatively generic variability, and is often further boosted by randomization and spatiotemporal disorder. By contrast, jointly achieving strong integration, preserved segregation, and elevated spatial and temporal clustering is far more demanding, and is not yet fully reproduced by any model considered here. In this respect, empirical dFC differs fundamentally from generic random variability. Its fluctuations are spatiotemporally organized, combining features reminiscent of heterogeneous, almost scale-free backbones with local cohesion and clustering more suggestive of small-world structure, without collapsing onto either template. Staticized surrogates make this tension visible but remain imposed by construction: \texttt{erand\_sfs} captures some heterogeneous backbone-like properties yet shows lower net fluidity and longer return latency, whereas \texttt{erand\_sws} captures local cohesion but exhibits excessive clustering, excessive retention, and slightly poorer irrigation. By contrast, some of these same features emerge mechanistically in the connectome-based \texttt{connMF} model operating in a regime of ``critical roaming'' between ignition and flaring transition lines \citep{Pathak2025}. That a mean-field model tuned independently of the present analyses can recover several distinct properties of dFC supports the plausibility of proximity to criticality as a major ingredient of the empirical dynamics. Yet the agreement remains more qualitative than quantitative (cf. Fig.~\ref{fig5} showing that the \texttt{connMF} model is indeed one of the ``top'' performers, but not the absolute quantitative best): simulated resting-state dFC is still offset from the empirical one and remains more temporally unstructured, consistent with its stronger irrigation but weaker temporal clustering. A closer quantitative match will likely require future whole-brain models to move beyond reproducing generic distributions of network fluidity and sequential variability, and instead capture the higher-order scaffold of dFC, including edge-based functional connectivity and the organization of meta-hubs controlling distributed stars of incident dynamic links \citep{Faskowitz2019, Lombardo2020,Arbabyazd2023}.

A further possibility is that dFC is shaped not only by constrained stochasticity, as in graph-based nulls, or by self-organized collective dynamics, as in connectome-based models, but also by ongoing computation itself. The fact that structured dFC supports a better integration/segregation tradeoff than null models does not imply that cognition simply ``runs on FC graphs''. A more plausible interpretation is that self-organized dynamics generates a baseline low-cost communication scaffold, onto which ongoing computations transiently imprint their own fingerprints. This view is consistent with the broader idea that resting-state dynamics provides a structured flow that task-related processing selectively modulates \citep{Cole2014}. It also yields a concrete prediction: the metrics introduced here ---cost-dependent spread, temporal clustering of strong links, and the segregation-integration balance---should vary systematically with task demands, arousal, engagement, and pathology in ways that simpler graph-theoretical null models or pure ``non-cognitive'' dynamic models may miss. In this sense, the present study does more than suggest that dFC can be useful. It points to spatiotemporally structured intermittency of communication as a candidate functional principle: neither static wiring nor random fluctuation, but a temporally organized communication geometry poised between cohesion and dispersion, between self-organization and control, and perhaps between the brain’s energetic constraints and its need for flexible computation.

\section{Methods}

\subsection{fMRI dataset}

We analyzed resting-state functional MRI (rs-fMRI) data from the test--retest dataset reported by \citet{termenon2016reliability}. This dataset comprises 100 healthy adults drawn from the Human Connectome Project, each scanned twice on separate days, yielding 200 rs-fMRI recordings in total. Because the acquisitions were obtained under resting conditions, the resulting time series reflect spontaneous fluctuations in brain activity rather than task-evoked responses.

Images were acquired on a customized 3T Siemens Connectome Skyra system using a multiband gradient-echo echo-planar imaging sequence with 2~mm isotropic resolution, repetition time $\mathrm{TR}=720$~ms, echo time $\mathrm{TE}=33.1$~ms, and multiband factor 8. Each run comprised 1,200 volumes, corresponding to approximately 14~min 24~s of rs-fMRI data.

For the present analysis, preprocessed data were represented at the regional level using the Automated Anatomical Labeling (AAL) atlas, yielding $N=89$ regions of interest per recording. Anatomical definitions therefore followed the standard AAL parcellation~\citep{tzourio2002automated}. The list of regions is provided in Table~\ref{tab:aal_regions}.

\subsection{From fMRI time series to temporal networks}

Each rs-fMRI recording was represented as a time-varying network of functional associations. Let
\begin{equation}
X \in \mathbb{R}^{T_0 \times N}
\end{equation}
denote the regional time series for one recording, where $T_0$ is the number of sampled time points and $N$ the number of brain regions.

Time-resolved functional connectivity was estimated using a sliding-window procedure. For window length $W$ and step size $\ell$, the connectivity matrix associated with window position $s$ was defined as
\begin{equation}
C^{(s)} = \mathrm{corr}\!\left(X_{[\,s:s+W-1\,]}\right),
\end{equation}
yielding a sequence of weighted symmetric matrices
\begin{equation}
\mathcal{C}=\{C^{(1)},C^{(2)},\dots,C^{(S)}\}, \qquad C^{(t)} \in [-1,1]^{N\times N},
\end{equation}
with
\begin{equation}
S = \left\lfloor \frac{T_0-W}{\ell} \right\rfloor + 1.
\end{equation}
Each matrix $C^{(t)}$ captures the pattern of statistical co-fluctuation among brain regions within one temporal window.

In all analyses, we used a window length of $W=20$ TR and unit step size $\ell=1$ TR. This choice yields a maximally overlapping sequence of connectivity frames, providing high temporal resolution while retaining a sufficient number of samples per window for stable correlation estimates.

To obtain binary temporal networks suitable for temporal-graph analysis, each weighted frame was thresholded so as to match a prescribed target mean degree. The binary adjacency matrix at frame $t$ was defined as
\begin{equation}
A_{ij}(t)=\mathbf{1}\{C_{ij}^{(t)}>\theta\}, \qquad i\neq j,
\end{equation}
with diagonal entries set to one,
\begin{equation}
A_{ii}(t)=1,
\end{equation}
so that walkers were allowed to remain at their current node while waiting for future links. The resulting temporal network is
\begin{equation}
\mathcal{T}=\{A(1),A(2),\dots,A(S)\}, \qquad A(t)\in\{0,1\}^{N\times N}.
\end{equation}

The threshold $\theta$ was selected such that the empirical mean degree approximately matched a prescribed target value $d_{\mathrm{target}}$. Writing
\begin{equation}
k_i(t)=\sum_{j\neq i}A_{ij}(t),
\end{equation}
the average degree across nodes and frames is
\begin{equation}
\bar d = \frac{1}{SN}\sum_{t=1}^{S}\sum_{i=1}^{N}k_i(t),
\end{equation}
and $\theta$ was chosen such that $\bar d \approx d_{\mathrm{target}}$.

This degree-controlled binarization defines the effective cost regime studied throughout the manuscript. Since cost is associated with keeping a link active for one time step, lower target degree corresponds to a higher-cost regime with sparser instantaneous frames, whereas higher target degree corresponds to a lower-cost regime with denser frames.

Before null-model generation, nodes showing no off-diagonal interaction throughout the full recording were removed from the active core network. This avoided artificially inflating measures based on unreachable pairs and reduced computational cost. When needed, trimmed nodes were reinserted afterwards as isolated rows and columns so that the final temporal networks remained aligned with the original atlas indexing.

\subsection{Temporal null models}

To determine which features of the temporal-network organization were specific to the empirical rs-fMRI data, we compared each observed temporal network with 15 null or control constructions, yielding 16 model classes in total including the empirical reference. Each model preserves selected coarse properties of the empirical sequence while randomizing others. Throughout, let
\begin{equation}
m_t = \sum_{1\leq i<j\leq N} A_{ij}(t)
\end{equation}
denote the number of undirected edges in frame $t$, and
\begin{equation}
\bar m = \frac{1}{S}\sum_{t=1}^{S}m_t
\end{equation}
the mean number of undirected edges per frame.

Null models were generated on the trimmed active core network and padded back to the original atlas size when needed.

\subsubsection{Empirical reference}

\paragraph{(1) \normalfont \texttt{fMRI\_tnet}}
The empirical temporal network itself, obtained from the rs-fMRI time series through sliding-window correlation followed by global thresholding to the prescribed target degree. This model preserves the observed temporal ordering, framewise density fluctuations, and detailed history of edge co-fluctuations, and serves as the reference condition throughout.

\subsubsection{Static-reference models}

\paragraph{(2) \normalfont\texttt{static\_fMRI}}
A static temporal network obtained by averaging the dynamic functional-connectivity matrices over time, ranking the off-diagonal entries of the resulting mean matrix, retaining the number of undirected edges implied by the target degree, and repeating the resulting binary adjacency across all $S$ frames. This removes temporal variability while preserving a static backbone derived from the same sliding-window FC representation.

\paragraph{(3) \normalfont\texttt{static\_er}}
A single Erd\H{o}s--R\'enyi-like graph with approximately $\bar m$ undirected edges, repeated across all frames.

\paragraph{(4) \normalfont\texttt{static\_sf}}
A single scale-free-like graph with approximately $\bar m$ undirected edges, repeated across all frames.

\paragraph{(5) \normalfont\texttt{static\_sw}}
A single small-world-like graph with approximately $\bar m$ undirected edges, repeated across all frames.

These three models preserve network size and mean density while eliminating empirical topology and temporal variation.

\subsubsection{Temporal-order control}

\paragraph{(6) \normalfont\texttt{time\_shuffle}}
The empirical frames are randomly permuted,
\begin{equation}
\widetilde{A}(t)=A\big(\pi(t)\big),
\end{equation}
where $\pi$ is a permutation of $\{1,\dots,S\}$. This preserves the full set of empirical frames, including their edge counts and within-frame topology, while destroying temporal ordering.

\subsubsection{Density-regularized model}

\paragraph{(7) \normalfont\texttt{uniform}}
A temporal network in which the mean number of undirected edges per frame is preserved but temporal fluctuations in frame density are suppressed. A new sequence of edge counts is generated so that its mean matches $\bar m$ and its temporal standard deviation is approximately zero, after which each frame is generated independently as an Erd\H{o}s--R\'enyi-like graph with the assigned number of edges.

\subsubsection{Edge-activation control}

\paragraph{(8) \normalfont\texttt{eAct}}
For each undirected edge $(i,j)$, the total number of activations across time,
\begin{equation}
a_{ij}=\sum_{t=1}^{S}A_{ij}(t),
\end{equation}
is preserved exactly, while the time points at which those activations occur are randomized. Thus, each edge remains active for the same number of frames as in the empirical network, but temporal coordination across edges is destroyed. In general, this model does not preserve the original framewise edge counts $\{m_t\}$.

\subsubsection{Template-based models}

The following models preserve the empirical framewise edge counts $\{m_t\}$ while imposing a global priority structure over edge recruitment.

\paragraph{(9) \normalfont\texttt{eActTemplate}}
Edges are ranked in descending order of empirical activation count $a_{ij}$. For a frame requiring $m_t$ edges, the initial edge set consists of the top $m_t$ ranked edges. To avoid a fully deterministic construction, a fraction of 5\% of these edges is then rewired independently at each frame. Specifically, $r=\lfloor 0.05\,m_t\rfloor$ selected edges are replaced by edges sampled from the complement set with rank-dependent probability
\begin{equation}
w_e \propto \frac{1}{(\mathrm{rank}(e)+1)^{\alpha}},
\end{equation}
with $\alpha=1$. When possible, edges that were never active in the empirical data are excluded from the sampling pool.

\paragraph{(10) \normalfont\texttt{eRndTemplate}}
Constructed identically to \texttt{eActTemplate}, including the same rewiring fraction, but using a random edge ranking instead of the empirical one. Any persistence in this model therefore arises from the template mechanism itself rather than from empirical activation statistics.

\subsubsection{Mechanistic generative model}

\paragraph{(11)\normalfont  \texttt{connMF}}
A temporal network derived from a mechanistic simulation of resting-state dynamics in a connectome-constrained mean-field model~\citep{Pathak2025}. Simulated neural activity was transformed into synthetic BOLD signals, from which a dFC sequence was extracted using the same sliding-window procedure as for the empirical data. The resulting weighted dFC stream was then binarized using the same thresholding pipeline. Unlike the purely statistical nulls, \texttt{connMF} does not explicitly preserve temporal ordering, edge-activation totals, or empirical framewise density fluctuations; rather, its temporal structure emerges through self-organized dynamics constrained by anatomical connectivity. As discussed by ~\citet{Pathak2025}, a different personalized mean-field model was fitted for every corresponding empirical resting state session.

\subsubsection{Frame-level topology controls}

These models preserve the empirical number of undirected edges in each frame while regenerating the adjacency pattern according to a prescribed graph family.

\paragraph{(12) \normalfont\texttt{erand\_er}}
At each frame $t$, a new random symmetric graph is generated independently with edge count matched to $m_t$.

\paragraph{(13) \normalfont\texttt{erand\_sws}}
Each frame is generated independently as a small-world-like graph with edge count matched to $m_t$, while preserving node labels across frames.

\paragraph{(14) \normalfont\texttt{erand\_sfs}}
Analogous to \texttt{erand\_sws}, but with each frame generated as a scale-free-like graph.

\paragraph{(15) \normalfont\texttt{erand\_swr}}
A relabeled version of \texttt{erand\_sws}, obtained by applying an independent random permutation of node labels at each frame. This preserves framewise edge count and small-world-like topology while destroying node-specific temporal persistence.

\paragraph{(16) \normalfont\texttt{erand\_sfr}}
The relabeled analogue of \texttt{erand\_sfs}, constructed by combining scale-free-like frame generation with independent random permutation of node labels at each frame.
\\

Whenever random, small-world, or scale-free graph families were required, we used standard procedures to generate them as described in previous work ~\cite{mengiste2015effect}.

\subsection{Temporal-network measures}

For each temporal network we extracted measures of integration, recurrence, local cohesion, memory, and temporal reconfiguration. The first four measures were computed under two propagation regimes, termed \emph{smart} and \emph{drunk}.

\subsubsection{Temporal paths and propagation regimes}

All integration measures were based on time-respecting paths. A temporal path from node $i$ to node $j$ is a sequence
\begin{equation}
(i=v_0,v_1,\dots,v_m=j)
\end{equation}
with associated nondecreasing frame indices
\begin{equation}
1\leq t_1\leq t_2\leq \dots \leq t_m\leq S,
\end{equation}
such that
\begin{equation}
A_{v_{r-1}v_r}(t_r)=1
\end{equation}
for every step $r$. Because $A_{ii}(t)=1$, a path may also remain at a node while waiting for future connections.

In the \emph{smart} regime, propagation follows an earliest-arrival strategy and therefore captures deterministic temporal shortest paths. In the \emph{drunk} regime, walkers move stochastically among currently available neighbors, thereby providing a random-walk measure of temporal accessibility.

\subsubsection{Integration measures under smart and drunk propagation}

For each ordered node pair $(i,j)$, let $d^{\mathrm{smart}}_{ij}$ denote the earliest-arrival temporal distance under smart propagation and $d^{\mathrm{drunk}}_{ij}$ the first-passage-time estimate under drunk propagation. For either regime, we computed the following measures.

\paragraph{\normalfont Irrigation}
Irrigation (also irrigation reach) is the number of ordered node pairs connected by at least one time-respecting path within the observation horizon.
\begin{equation}
\mathrm{Irrigation} = \sum_{i\neq j}\mathbf{1}\{d_{ij}<\infty\}.
\end{equation}

\paragraph{Latency}
Latency measures the mean temporal distance among reachable pairs.
\begin{equation}
\mathrm{Latency}=
\frac{\sum_{i\neq j} d_{ij}\,\mathbf{1}\{d_{ij}<\infty\}}
{\sum_{i\neq j}\mathbf{1}\{d_{ij}<\infty\}}.
\end{equation}

\paragraph{Resistance}
Resistance (also irrigation resistance) incorporates both finite path length and outright inaccessibility. If $d_{\max}$ denotes the maximum finite temporal distance observed in the network, unreachable pairs are assigned penalty $d_{\max}+1$:
\begin{equation}
\mathrm{Resistance} = \frac{1}{N(N-1)}\sum_{i\neq j}\widetilde d_{ij},
\end{equation}
where
\begin{equation}
\widetilde d_{ij}=
\begin{cases}
d_{ij}, & d_{ij}<\infty,\\
d_{\max}+1, & d_{ij}=\infty.
\end{cases}
\end{equation}

\paragraph{Inaccessibility}
Inaccessibility counts the number of ordered node pairs that cannot be connected by any time-respecting path.
\begin{equation}
\mathrm{Inaccessibility} = \sum_{i\neq j}\mathbf{1}\{d_{ij}=\infty\}.
\end{equation}

\subsubsection{Segregation measures}

\paragraph{Return latency}

Return latency quantifies how rapidly a node can be reached again after departure under deterministic time-respecting propagation. For each node and admissible start time, we tracked the first elapsed number of frames for which a nontrivial time-respecting path beginning at that node returned to it. Such first return time, excluding self-loops, can be semi-analytically estimated from the knowledge of the temporal network stream of frames $A(t)$. Considering a node $i$ as source and defining a $i$-punched frame as:
\begin{equation}
A^{(i)}_{kl}(t) = 
\begin{cases}
A_{kl}(t), & k, l \neq i,\\
0, & k = l = i.
\end{cases}
\end{equation}
and define recursively the sequential $i$-punched product of time-resolved adjacency matrices as :
\begin{equation}
\begin{array}{cc}
     \Pi^{(i)}(t,K) &  = \Pi^{(i)}(t, K-1) \cdot A^{(i)}(t+K)\\
     \Pi^{(i)}(t,1)& =  A^{(i)}(t) \cdot A^{(i)}(t+1)
\end{array}
\end{equation}
then the return latency at time $t$ for source node $i$ is nothing else than:
\begin{equation}
L_i^{\mathrm{return}}(t)=\min\left\{K\ge 1:\; \Pi^{(i)}_{ii}(t,K)>0\right\}    
\end{equation}
Return latencies were averaged across admissible start times $t$ at the node level and summarized at the network level by the median node-wise return latency. Lower values indicate faster local recurrence.

\paragraph{Retention}

Retention quantifies the tendency of nodes to reconnect to previously encountered neighbors. Let $\mathcal{N}_i(t)$ denote the set of neighbors of node $i$ at frame $t$, excluding self-connections, and let
\begin{equation}
\mathcal{P}_i(t)=\bigcup_{u<t}\mathcal{N}_i(u)
\end{equation}
be the set of neighbors previously encountered by node $i$. Instantaneous retention is defined as
\begin{equation}
R_i(t)=\frac{|\mathcal{N}_i(t)\cap \mathcal{P}_i(t)|}{|\mathcal{N}_i(t)|},
\qquad \text{for } \mathcal{N}_i(t)\neq \varnothing.
\end{equation}
Node-wise retention was obtained by averaging $R_i(t)$ across frames, and network-level retention by taking the median across nodes. High retention indicates preferential reconnection to previously encountered partners.

\paragraph{Static clustering}

For each frame $t$ and node $i$, the standard clustering coefficient was computed from the instantaneous neighborhood:
\begin{equation}
C_i(t)=
\frac{\#\{\text{links among neighbors of }i\text{ at }t\}}
{\binom{k_i(t)}{2}},
\end{equation}
with $C_i(t)=0$ whenever $k_i(t)<2$. Static clustering was averaged over time at the node level and summarized by the median across nodes.

\paragraph{Temporal clustering}

Temporal clustering extends triangle closure across adjacent frames. Suppose node $i$ is connected to node $j$ at frame $t_1$ and to node $k$ at frame
\begin{equation}
t_2=t_1+\alpha, \qquad 0<\alpha\leq \Delta.
\end{equation}
A temporal triangle centered on $i$ is completed if nodes $j$ and $k$ are also connected at some intermediate or matching frame
\begin{equation}
t_3=t_1+\beta, \qquad 0<\beta\leq \alpha,
\end{equation}
such that
\begin{equation}
t_1 < t_3 \leq t_2 \leq t_1+\Delta.
\end{equation}
In the present analyses we used $\Delta=1$, enforcing a strictly local notion of temporal cohesion across consecutive frames. Node-wise temporal clustering coefficients were computed as the ratio of valid temporal triangle closures to admissible ordered triplets centered on that node, and then summarized by the median across nodes.

\subsubsection{Statism/dynamism measures}

\paragraph{Similarity}

Similarity between consecutive network states was quantified by the cosine similarity of their vectorized undirected edge sets. Let $\mathbf{a}^{(t)}\in\{0,1\}^E$ denote the flattened lower-triangular edge vector at frame $t$, where $E=\binom{N}{2}$. Then
\begin{equation}
\mathrm{sim}(t,t+1)=
\frac{\langle \mathbf{a}^{(t)},\mathbf{a}^{(t+1)}\rangle}
{\|\mathbf{a}^{(t)}\|_2\,\|\mathbf{a}^{(t+1)}\|_2},
\end{equation}
with standard conventions for zero vectors. The global similarity score is
\begin{equation}
\mathrm{Similarity}=\frac{1}{S-1}\sum_{t=1}^{S-1}\mathrm{sim}(t,t+1).
\end{equation}
High similarity indicates strong persistence of network structure across frames.

\paragraph{Net fluidity}

Net fluidity combines global temporal heterogeneity with frame-to-frame change. For each edge $e$, let $p_e$ denote its activation probability across frames. The corresponding entropy is
\begin{equation}
H_e = -\big[p_e\log p_e + (1-p_e)\log(1-p_e)\big],
\end{equation}
and the normalized global entropy is
\begin{equation}
\mathrm{gEnt}=\frac{\sum_{e=1}^{E}H_e}{E\log 2}.
\end{equation}
Net fluidity is then defined as
\begin{equation}
\mathrm{Net\ fluidity}=\mathrm{gEnt}\,\big(1-\mathrm{Similarity}\big).
\end{equation}
This quantity is high when edge activations are globally heterogeneous and consecutive frames differ strongly from one another.

\section*{Acknowledgements}

This work was supported by the PEPR Sant\'e Num\'erique program (France 2030), project ``Brain Health Trajectories (BHT)'', implemented by the Agence Nationale de la Recherche (ANR) under grant number ANR-22-PESN-0012-BHT. We wish to thank Alain Barrat, Caio Seguin and Sinisa Pajevic for inspiring discussions and Anagh Pathak for sharing time-series from connectome-based simulations.

\section*{Declarations}

\begin{itemize}
\item Funding: PEPR Sant\'e Num\'erique ``Brain Health Trajectory'' (ANR-22-PESN-0012-BHT). 
\item Competing interests: none
\item Ethics approval and consent to participate: Not applicable
\item Consent for publication: all authors have approved the current manuscript version.
\item Data availability: analyzed open data will be made available at publication acceptance and are available upon request. 
\item Material availability: Not applicable
\item Code availability: code downloadable at \href{https://github.com/absima/tnet}{github.com/absima/tnet}
\item Author contribution: SM and DB performed research, designed research, wrote the article.
\end{itemize}

\noindent

\bibliography{tnet-bibliography}

@article{Hutchison2013, 
year = {2013}, 
month = {10}, 
title = {{Dynamic functional connectivity: Promise, issues, and interpretations}}, 
author = {Hutchison, R. Matthew and Womelsdorf, Thilo and Allen, Elena A. and Bandettini, Peter A. and Calhoun, Vince D. and Corbetta, Maurizio and Penna, Stefania Della and Duyn, Jeff H. and Glover, Gary H. and Gonzalez-Castillo, Javier and Handwerker, Daniel A. and Keilholz, Shella and Kiviniemi, Vesa and Leopold, David A. and Pasquale, Francesco de and Sporns, Olaf and Walter, Martin and Chang, Catie}, 
journal = {NeuroImage}, 
issn = {1053-8119}, 
doi = {10.1016/j.neuroimage.2013.05.079}, 
pmid = {23707587}, 
pmcid = {PMC3807588}, 
pages = {360--378}, 
volume = {80}
}

@article{Calhoun2014, 
year = {2014}, 
month = {10}, 
title = {{The Chronnectome: Time-Varying Connectivity Networks as the Next Frontier in fMRI Data Discovery}}, 
author = {Calhoun, Vince D. and Miller, Robyn and Pearlson, Godfrey and Adalı, Tulay}, 
journal = {Neuron}, 
issn = {0896-6273}, 
doi = {10.1016/j.neuron.2014.10.015}, 
pmid = {25374354}, 
pmcid = {PMC4372723}, 
pages = {262--274}, 
number = {2}, 
volume = {84}
}

@article{Preti2017, 
year = {2017}, 
month = {10}, 
title = {{The dynamic functional connectome: State-of-the-art and perspectives}}, 
author = {Preti, Maria Giulia and Bolton, Thomas AW and Ville, Dimitri Van De}, 
journal = {NeuroImage}, 
issn = {1053-8119}, 
doi = {10.1016/j.neuroimage.2016.12.061}, 
pmid = {28034766}, 
pages = {41--54}, 
volume = {160}
}

@article{Pedreschi2020, 
year = {2020}, 
title = {{Dynamic core-periphery structure of information sharing networks in entorhinal cortex and hippocampus}}, 
author = {Pedreschi, Nicola and Bernard, Christophe and Clawson, Wesley and Quilichini, Pascale and Barrat, Alain and Battaglia, Demian}, 
journal = {Network Neuroscience}, 
doi = {10.1162/netn\_a\_00142}, 
pmid = {33615098}, 
pmcid = {PMC7888487}, 
pages = {946--975}, 
number = {3}, 
volume = {4}
}

@article{Pedreschi2022, 
year = {2022}, 
title = {{The temporal rich club phenomenon}}, 
author = {Pedreschi, Nicola and Battaglia, Demian and Barrat, Alain}, 
journal = {Nature Physics}, 
issn = {1745-2473}, 
doi = {10.1038/s41567-022-01634-8}, 
pages = {1--8}
}

@article{Arbabyazd2023, 
year = {2023}, 
title = {{State-switching and high-order spatiotemporal organization of dynamic functional connectivity are disrupted by Alzheimer’s disease}}, 
author = {Arbabyazd, Lucas and Petkoski, Spase and Breakspear, Michael and Solodkin, Ana and Battaglia, Demian and Jirsa, Viktor}, 
journal = {Network Neuroscience}, 
doi = {10.1162/netn\_a\_00332}, 
pmid = {38144688}, 
pmcid = {PMC10727776}, 
pages = {1420--1451}, 
number = {4}, 
volume = {7}
}

@article{Clawson2023, 
year = {2023}, 
title = {{Perturbed information processing complexity in experimental epilepsy}}, 
author = {Clawson, Wesley and Waked, Benjamin and Madec, Tanguy and Ghestem, Antoine and Quilichini, Pascale P and Battaglia, Demian and Bernard, Christophe}, 
journal = {The Journal of Neuroscience}, 
issn = {0270-6474}, 
doi = {10.1523/jneurosci.0383-23.2023}, 
pmid = {37550052}, 
pages = {JN--RM-0383-23}
}

@article{Pedreschi2026, 
year = {2026}, 
title = {{States of dynamic connectivity flow in temporal multiplex networks: a
case study in human epilepsy and postictal aphasia}}, 
author = {Pedreschi, Nicola and Trebuchon, Agnes and Barrat, Alain and Battaglia, Demian}, 
journal = {Network Neuroscience}, 
doi = {}, 
pmid = {???}, 
pmcid = {???}, 
pages = {}, 
number = {}, 
volume = {in press}
}

@article{Deco2013, 
year = {2013}, 
title = {{Resting brains never rest: computational insights into potential cognitive architectures.}}, 
author = {Deco, Gustavo and Jirsa, Viktor K and Mcintosh, Anthony R}, 
journal = {Trends in neurosciences}, 
issn = {0166-2236}, 
doi = {10.1016/j.tins.2013.03.001}, 
pmid = {23561718}, 
pages = {268 -- 274}, 
number = {5}, 
volume = {36}, 
language = {English}, 
month = {05}
}

@article{Hansen2015, 
year = {2015}, 
title = {{Functional connectivity dynamics: modeling the switching behavior of the resting state.}}, 
author = {Hansen, Enrique C A and Battaglia, Demian and Spiegler, Andreas and Deco, Gustavo and Jirsa, Viktor K}, 
journal = {NeuroImage}, 
issn = {1053-8119}, 
doi = {10.1016/j.neuroimage.2014.11.001}, 
pmid = {25462790}, 
pages = {525 535}, 
number = {Cereb. Cortex 2012}, 
volume = {105}, 
language = {English}, 
month = {01}
}

@article{Pathak2025, 
year = {2025}, 
month = {12}, 
title = {{The critical roaming hypothesis: arousal-driven transitions across critical lines reproduce human functional connectivity dynamics}}, 
author = {Pathak, Anagh and Battaglia, Demian}, 
journal = {bioRxiv}, 
doi = {10.64898/2025.12.29.696846}, 
pages = {2025.12.29.696846}
}

@article{Bassett2011, 
year = {2011}, 
title = {{Dynamic reconfiguration of human brain networks during learning.}}, 
author = {Bassett, Danielle S and Wymbs, Nicholas F and Porter, Mason A and Mucha, Peter J and Carlson, Jean M and Grafton, Scott T}, 
journal = {Proceedings of the National Academy of Sciences of the United States of America}, 
issn = {0027-8424}, 
doi = {10.1073/pnas.1018985108}, 
pmid = {21502525}, 
pmcid = {PMC3088578}, 
eprint = {1010.3775}, 
pages = {7641 7646}, 
number = {18}, 
volume = {108}, 
language = {English}, 
month = {05}
}

@article{Braun2015, 
year = {2015}, 
title = {{Dynamic reconfiguration of frontal brain networks during executive cognition in humans.}}, 
author = {Braun, Urs and Schäfer, Axel and Walter, Henrik and Erk, Susanne and Romanczuk-Seiferth, Nina and Haddad, Leila and Schweiger, Janina I and Grimm, Oliver and Heinz, Andreas and Tost, Heike and Meyer-Lindenberg, Andreas and Bassett, Danielle S}, 
journal = {Proceedings of the National Academy of Sciences of the United States of America}, 
issn = {0027-8424}, 
doi = {10.1073/pnas.1422487112}, 
pmid = {26324898}, 
pmcid = {PMC4577153}, 
pages = {11678 11683}, 
number = {37}, 
volume = {112}, 
language = {English}, 
month = {09}
}

@article{Shine2016, 
year = {2016}, 
title = {{The Dynamics of Functional Brain Networks: Integrated Network States during Cognitive Task Performance.}}, 
author = {Shine, James M and Bissett, Patrick G and Bell, Peter T and Koyejo, Oluwasanmi and Balsters, Joshua H and Gorgolewski, Krzysztof J and Moodie, Craig A and Poldrack, Russell A}, 
journal = {Neuron}, 
issn = {0896-6273}, 
doi = {10.1016/j.neuron.2016.09.018}, 
pmid = {27693256}, 
eprint = {1511.02976}, 
pages = {544 554}, 
number = {2}, 
volume = {92}, 
language = {English}, 
month = {10}
}

@article{Battaglia2020, 
year = {2020}, 
title = {{Dynamic Functional Connectivity between order and randomness and its evolution across the human adult lifespan}}, 
author = {Battaglia, Demian and Boudou, Thomas and Hansen, Enrique C.A. and Lombardo, Diego and Chettouf, Sabrina and Daffertshofer, Andreas and McIntosh, Anthony R. and Zimmermann, Joelle and Ritter, Petra and Jirsa, Viktor}, 
journal = {NeuroImage}, 
issn = {1053-8119}, 
doi = {10.1016/j.neuroimage.2020.117156}, 
pmid = {32698027}, 
pages = {117156}, 
volume = {222}
}

@article{Lombardo2020, 
year = {2020}, 
title = {{Modular slowing of resting-state dynamic functional connectivity as a marker of cognitive dysfunction induced by sleep deprivation}}, 
author = {Lombardo, Diego and Cassé-Perrot, Catherine and Ranjeva, Jean-Philippe and Troter, Arnaud Le and Guye, Maxime and Wirsich, Jonathan and Payoux, Pierre and Bartrés-Faz, David and Bordet, Régis and Richardson, Jill C and Felician, Olivier and Jirsa, Viktor and Blin, Olivier and Didic, Mira and Battaglia, Demian}, 
journal = {NeuroImage}, 
issn = {1053-8119}, 
doi = {10.1016/j.neuroimage.2020.117155}, 
pmid = {32736002}, 
pages = {117155}, 
volume = {222}
}

@article{Tononi1994, 
year = {1994}, 
title = {{A measure for brain complexity: relating functional segregation and integration in the nervous system.}}, 
author = {Tononi, G and Sporns, O and Edelman, G M}, 
journal = {Proceedings of the National Academy of Sciences}, 
issn = {0027-8424}, 
doi = {10.1073/pnas.91.11.5033}, 
pmid = {8197179}, 
pmcid = {PMC43925}, 
pages = {5033--5037}, 
number = {11}, 
volume = {91}
}

@article{Sporns2013, 
year = {2013}, 
month = {4}, 
title = {{Network attributes for segregation and integration in the human brain}}, 
author = {Sporns, Olaf}, 
journal = {Current Opinion in Neurobiology}, 
issn = {0959-4388}, 
doi = {10.1016/j.conb.2012.11.015}, 
pmid = {23294553}, 
pages = {162--171}, 
number = {2}, 
volume = {23}
}

@article{Cabral2017, 
year = {2017}, 
title = {{Functional connectivity dynamically evolves on multiple time-scales over a static structural connectome: Models and mechanisms}}, 
author = {Cabral, Joana and Kringelbach, Morten L and Deco, Gustavo}, 
journal = {NeuroImage}, 
issn = {1053-8119}, 
doi = {10.1016/j.neuroimage.2017.03.045}, 
pmid = {28343985}, 
pages = {84--96}, 
number = {Phys. Rev. E, State Phys., Plasmas, Fluids, Relat. Interdiscip. Top. 48 1993}, 
volume = {160}
}

@article{Shine2019, 
year = {2019}, 
title = {{Neuromodulatory Influences on Integration and Segregation in the Brain}}, 
author = {Shine, James M.}, 
journal = {Trends in Cognitive Sciences}, 
issn = {1364-6613}, 
doi = {10.1016/j.tics.2019.04.002}, 
pmid = {31076192}, 
pages = {572--583}, 
number = {7}, 
volume = {23}
}

@article{Deco2015, 
year = {2015}, 
title = {{Rethinking segregation and integration: contributions of whole-brain modelling.}}, 
author = {Deco, Gustavo and Tononi, Giulio and Boly, Melanie and Kringelbach, Morten L}, 
journal = {Nature Reviews Neuroscience}, 
issn = {1471-003x}, 
doi = {10.1038/nrn3963}, 
pmid = {26081790}, 
pages = {430 439}, 
number = {7}, 
volume = {16}, 
language = {English}
}

@article{Raichle2006, 
year = {2006}, 
month = {11}, 
title = {{The Brain's Dark Energy}}, 
author = {Raichle, Marcus E}, 
journal = {Science}, 
issn = {0036-8075}, 
doi = {10.1126/science.1134405}, 
pmid = {17124311}, 
pages = {1249--1250}, 
number = {5803}, 
volume = {314}
}

@article{Tomasi2013, 
year = {2013}, 
month = {8}, 
title = {{Energetic cost of brain functional connectivity}}, 
author = {Tomasi, Dardo and Wang, Gene-Jack and Volkow, Nora D.}, 
journal = {Proceedings of the National Academy of Sciences}, 
issn = {0027-8424}, 
doi = {10.1073/pnas.1303346110}, 
pmid = {23898179}, 
pmcid = {PMC3746878}, 
pages = {13642--13647}, 
number = {33}, 
volume = {110}
}

@article{Volpi2024, 
year = {2024}, 
month = {2}, 
title = {{The brain’s “dark energy” puzzle: How strongly is glucose metabolism linked to resting-state brain activity?}}, 
author = {Volpi, Tommaso and Silvestri, Erica and Aiello, Marco and Lee, John J and Vlassenko, Andrei G and Goyal, Manu S and Corbetta, Maurizio and Bertoldo, Alessandra}, 
journal = {Journal of Cerebral Blood Flow \& Metabolism}, 
issn = {0271-678X}, 
doi = {10.1177/0271678X241237974}, 
pmid = {38443762}, 
pmcid = {PMC11342718}, 
pages = {1433--1449}, 
number = {8}, 
volume = {44}
}

@article{Kivela2012, 
year = {2012}, 
month = {3}, 
title = {{Multiscale analysis of spreading in a large communication network}}, 
author = {Kivelä, Mikko and Pan, Raj Kumar and Kaski, Kimmo and Kertész, János and Saramäki, Jari and Karsai, Márton}, 
journal = {Journal of Statistical Mechanics: Theory and Experiment}, 
doi = {10.1088/1742-5468/2012/03/P03005}, 
eprint = {1112.4312}, 
pages = {P03005}, 
number = {03}, 
volume = {2012}
}

@article{Gauvin2013, 
year = {2013}, 
title = {{Activity clocks: spreading dynamics on temporal networks of human contact}}, 
author = {Gauvin, Laetitia and Panisson, André and Cattuto, Ciro and Barrat, Alain}, 
journal = {Scientific Reports}, 
doi = {10.1038/srep03099}, 
pmid = {24172876}, 
pmcid = {PMC3813939}, 
eprint = {1306.4626}, 
pages = {3099}, 
number = {1}, 
volume = {3}
}

@article{Holme2012, 
year = {2012}, 
title = {{Temporal networks}}, 
author = {Holme, Petter and Saramäki, Jari}, 
journal = {Physics reports}, 
issn = {0370-1573}, 
doi = {10.1016/j.physrep.2012.03.001}, 
eprint = {1108.1780}, 
pages = {97 125}, 
number = {3}, 
volume = {519}, 
language = {English}
}

@article{Kivela2014, 
year = {2014}, 
title = {{Multilayer networks}}, 
author = {Kivelä, M and Arenas, A and Barthelemy, M}, 
journal = {Journal of Complex Networks}, 
issn = {2051-1310}, 
doi = {10.1093/comnet/cnu016}, 
eprint = {1309.7233}, 
pages = {203 271}, 
number = {3}, 
volume = {2}, 
language = {English}
}

@article{Shannon1948, 
year = {1948}, 
month = {7}, 
title = {{A Mathematical Theory of Communication}}, 
author = {Shannon, C E}, 
journal = {Bell System Technical Journal}, 
issn = {0005-8580}, 
doi = {10.1002/j.1538-7305.1948.tb01338.x}, 
pages = {379--423}, 
number = {3}, 
volume = {27}
}

@article{Grothe2018, 
year = {2018}, 
month = {4}, 
title = {{Attention Selectively Gates Afferent Signal Transmission to Area V4}}, 
author = {Grothe, Iris and Rotermund, David and Neitzel, Simon David and Mandon, Sunita and Ernst, Udo Alexander and Kreiter, Andreas K. and Pawelzik, Klaus Richard}, 
journal = {The Journal of Neuroscience}, 
issn = {0270-6474}, 
doi = {10.1523/JNEUROSCI.2221-17.2018}, 
pmid = {29618546}, 
pmcid = {PMC6596051}, 
pages = {3441--3452}, 
number = {14}, 
volume = {38}
}

@article{Papadopoulos2020, 
year = {2020}, 
title = {{Relations between large-scale brain connectivity and effects of regional stimulation depend on collective dynamical state}}, 
author = {Papadopoulos, Lia and Lynn, Christopher W. and Battaglia, Demian and Bassett, Danielle S.}, 
journal = {PLoS Computational Biology}, 
issn = {1553-734X}, 
doi = {10.1371/journal.pcbi.1008144}, 
pmid = {32886673}, 
pmcid = {PMC7537889}, 
pages = {e1008144}, 
number = {9}, 
volume = {16}
}

@article{Kempe2002, 
year = {2002}, 
month = {6}, 
title = {{Connectivity and Inference Problems for Temporal Networks}}, 
author = {Kempe, David and Kleinberg, Jon and Kumar, Amit}, 
journal = {Journal of Computer and System Sciences}, 
issn = {0022-0000}, 
doi = {10.1006/jcss.2002.1829}, 
pages = {820--842}, 
number = {4}, 
volume = {64}
}

@article{Starnini2013, 
year = {2013}, 
month = {11}, 
title = {{Immunization strategies for epidemic processes in time-varying contact networks}}, 
author = {Starnini, Michele and Machens, Anna and Cattuto, Ciro and Barrat, Alain and Pastor-Satorras, Romualdo}, 
journal = {Journal of Theoretical Biology}, 
issn = {0022-5193}, 
doi = {10.1016/j.jtbi.2013.07.004}, 
pmid = {23871715}, 
eprint = {1305.2357}, 
pages = {89--100}, 
volume = {337}
}

@article{Valdano2015, 
year = {2015}, 
month = {1}, 
title = {{Analytical Computation of the Epidemic Threshold on Temporal Networks}}, 
author = {Valdano, Eugenio and Ferreri, Luca and Poletto, Chiara and Colizza, Vittoria}, 
journal = {Physical Review X}, 
doi = {10.1103/PhysRevX.5.021005}, 
pages = {021005}, 
number = {2}, 
volume = {5}
}

@article{ER1960, 
year = {1960}, 
month = {}, 
title = {{On the evolution of random graphs}}, 
author = {Erdös, P. and Rényi, A.}, 
journal = {Publications of the Mathematical Institute of the Hungarian Academy of Sciences}, 
doi = {}, 
pages = {17--61}, 
number = {}, 
volume = {5}
}

@article{Cardillo2014, 
year = {2014}, 
month = {1}, 
title = {{Evolutionary dynamics of time-resolved social interactions}}, 
author = {Cardillo, Alessio and Petri, Giovanni and Nicosia, Vincenzo and Sinatra, Roberta and Gómez-Gardeñes, Jesús and Latora, Vito}, 
journal = {Physical Review E}, 
issn = {1539-3755}, 
doi = {10.1103/PhysRevE.90.052825}, 
pmid = {25493851}, 
eprint = {1302.0558}, 
pages = {052825}, 
number = {5}, 
volume = {90}
}

@article{Cozzo2015, 
year = {2015}, 
month = {7}, 
title = {{Structure of triadic relations in multiplex networks}}, 
author = {Cozzo, Emanuele and Kivelä, Mikko and Domenico, Manlio De and Solé-Ribalta, Albert and Arenas, Alex and Gómez, Sergio and Porter, Mason A and Moreno, Yamir}, 
journal = {New Journal of Physics}, 
doi = {10.1088/1367-2630/17/7/073029}, 
eprint = {1307.6780}, 
pages = {073029}, 
number = {7}, 
volume = {17}
}

@article{Bail2024, 
year = {2024}, 
month = {11}, 
title = {{Flow of temporal network properties under local aggregation and time shuffling: a tool for characterizing, comparing and classifying temporal networks}}, 
author = {Bail, Didier Le and Génois, Mathieu and Barrat, Alain}, 
journal = {Journal of Physics A: Mathematical and Theoretical}, 
issn = {1751-8113}, 
doi = {10.1088/1751-8121/ad7b8e}, 
eprint = {2310.09112}, 
pages = {435002}, 
number = {43}, 
volume = {57}
}

@article{Tang2009, 
year = {2009}, 
month = {8}, 
title = {{Temporal distance metrics for social network analysis}}, 
author = {Tang, John and Musolesi, Mirco and Mascolo, Cecilia and Latora, Vito}, 
journal = {Proceedings of the 2nd ACM workshop on Online social networks}, 
doi = {10.1145/1592665.1592674}, 
pages = {31--36}
}

@article{Masuda2013, 
year = {2013}, 
month = {1}, 
title = {{Temporal Networks: Slowing Down Diffusion by Long Lasting Interactions}}, 
author = {Masuda, Naoki and Klemm, Konstantin and Eguíluz, Víctor M.}, 
journal = {Physical Review Letters}, 
issn = {0031-9007}, 
doi = {10.1103/PhysRevLett.111.188701}, 
pmid = {24237569}, 
eprint = {1305.2938}, 
pages = {188701}, 
number = {18}, 
volume = {111}
}

@article{Scholtes2014, 
year = {2014}, 
month = {9}, 
title = {{Causality-driven slow-down and speed-up of diffusion in non-Markovian temporal networks}}, 
author = {Scholtes, Ingo and Wider, Nicolas and Pfitzner, René and Garas, Antonios and Tessone, Claudio J. and Schweitzer, Frank}, 
journal = {Nature Communications}, 
doi = {10.1038/ncomms6024}, 
pmid = {25248462}, 
eprint = {1307.4030}, 
pages = {5024}, 
number = {1}, 
volume = {5}
}

@article{Delvenne2015, 
year = {2015}, 
month = {6}, 
title = {{Diffusion on networked systems is a question of time or structure}}, 
author = {Delvenne, Jean-Charles and Lambiotte, Renaud and Rocha, Luis E. C.}, 
journal = {Nature Communications}, 
doi = {10.1038/ncomms8366}, 
pmid = {26054307}, 
eprint = {1309.4155}, 
pages = {7366}, 
number = {1}, 
volume = {6}
}

@article{Li2017, 
year = {2017}, 
title = {{The fundamental advantages of temporal networks}}, 
author = {Li, A. and Cornelius, S. P. and Liu, Y.-Y. and Wang, L. and Barabási, A.-L.}, 
journal = {Science}, 
issn = {0036-8075}, 
doi = {10.1126/science.aai7488}, 
pmid = {29170233}, 
eprint = {1607.06168}, 
pages = {1042--1046}, 
number = {6366}, 
volume = {358}
}

@article{Pajevic2012, 
year = {2012}, 
title = {{The organization of strong links in complex networks}}, 
author = {Pajevic, Sinisa and Plenz, Dietmar}, 
journal = {Nature Physics}, 
issn = {1745-2481}, 
doi = {10.1038/nphys2257}, 
pmid = {28890731}, 
eprint = {1109.2577}, 
pages = {429 -- 436}, 
number = {5}, 
volume = {8}, 
month = {03}
}

@article{Seguin2022, 
year = {2022}, 
month = {8}, 
title = {{Network communication models narrow the gap between the modular organization of structural and functional brain networks}}, 
author = {Seguin, Caio and L, Sina Mansour and Sporns, Olaf and Zalesky, Andrew and Calamante, Fernando}, 
journal = {NeuroImage}, 
issn = {1053-8119}, 
doi = {10.1016/j.neuroimage.2022.119323}, 
pmid = {35605765}, 
pages = {119323}, 
volume = {257}
}

@article{Goni2014, 
year = {2014}, 
month = {1}, 
title = {{Resting-brain functional connectivity predicted by analytic measures of network communication}}, 
author = {Goñi, Joaquín and Heuvel, Martijn P. van den and Avena-Koenigsberger, Andrea and Mendizabal, Nieves Velez de and Betzel, Richard F. and Griffa, Alessandra and Hagmann, Patric and Corominas-Murtra, Bernat and Thiran, Jean-Philippe and Sporns, Olaf}, 
journal = {Proceedings of the National Academy of Sciences}, 
issn = {0027-8424}, 
doi = {10.1073/pnas.1315529111}, 
pmid = {24379387}, 
pmcid = {PMC3896172}, 
pages = {833--838}, 
number = {2}, 
volume = {111}
}

@article{Seguin2023, 
year = {2023}, 
title = {{Brain network communication: concepts, models and applications}}, 
author = {Seguin, Caio and Sporns, Olaf and Zalesky, Andrew}, 
journal = {Nature Reviews Neuroscience}, 
issn = {1471-003X}, 
doi = {10.1038/s41583-023-00718-5}, 
pmid = {37438433}, 
pages = {557--574}, 
number = {9}, 
volume = {24}
}

@article{Avena2018, 
year = {2018}, 
month = {1}, 
title = {{Communication dynamics in complex brain networks}}, 
author = {Avena-Koenigsberger, Andrea and Misic, Bratislav and Sporns, Olaf}, 
journal = {Nature Reviews Neuroscience}, 
issn = {1471-003X}, 
doi = {10.1038/nrn.2017.149}, 
pmid = {29238085}, 
pages = {17--33}, 
number = {1}, 
volume = {19}
}

@article{Battaglia2012, 
year = {2012}, 
title = {{Dynamic Effective Connectivity of Inter-Areal Brain Circuits}}, 
author = {Battaglia, Demian and Witt, Annette and Wolf, Fred and Geisel, Theo}, 
journal = {PLoS Computational Biology}, 
issn = {1553-734X}, 
doi = {10.1371/journal.pcbi.1002438}, 
pmid = {22457614}, 
pmcid = {PMC3310731}, 
eprint = {1112.3968}, 
pages = {e1002438}, 
number = {3}, 
volume = {8}, 
language = {English}, 
month = {03}
}

@article{Kirst2016, 
year = {2016}, 
title = {{Dynamic information routing in complex networks.}}, 
author = {Kirst, Christoph and Timme, Marc and Battaglia, Demian}, 
journal = {Nature communications}, 
issn = {2041-1723}, 
doi = {10.1038/ncomms11061}, 
pmid = {27067257}, 
pmcid = {PMC4832059}, 
eprint = {1510.05033}, 
pages = {11061}, 
number = {1}, 
volume = {7}, 
language = {English}
}

@article{Vicente2008, 
year = {2008}, 
month = {11}, 
title = {{Dynamical relaying can yield zero time lag neuronal synchrony despite long conduction delays}}, 
author = {Vicente, Raul and Gollo, Leonardo L. and Mirasso, Claudio R. and Fischer, Ingo and Pipa, Gordon}, 
journal = {Proceedings of the National Academy of Sciences}, 
issn = {0027-8424}, 
doi = {10.1073/pnas.0809353105}, 
pmid = {18957544}, 
pmcid = {PMC2575223}, 
pages = {17157--17162}, 
number = {44}, 
volume = {105}
}

@article{Hlinka2011, 
year = {2011}, 
month = {2}, 
title = {{Functional connectivity in resting-state fMRI: Is linear correlation sufficient?}}, 
author = {Hlinka, Jaroslav and Paluš, Milan and Vejmelka, Martin and Mantini, Dante and Corbetta, Maurizio}, 
journal = {NeuroImage}, 
issn = {1053-8119}, 
doi = {10.1016/j.neuroimage.2010.08.042}, 
pmid = {20800096}, 
pmcid = {PMC4139498}, 
pages = {2218--2225}, 
number = {3}, 
volume = {54}
}

@book{ShannonWeaver1949,
  author    = {Shannon, Claude E. and Weaver, Warren},
  title     = {The Mathematical Theory of Communication},
  year      = {1949},
  publisher = {University of Illinois Press},
  address   = {Urbana},
  language  = {English}
}

@article{Seguin2023stim, 
year = {2023}, 
month = {5}, 
title = {{Communication dynamics in the human connectome shape the cortex-wide propagation of direct electrical stimulation}}, 
author = {Seguin, Caio and Jedynak, Maciej and David, Olivier and Mansour, Sina and Sporns, Olaf and Zalesky, Andrew}, 
journal = {Neuron}, 
issn = {0896-6273}, 
doi = {10.1016/j.neuron.2023.01.027}, 
pmid = {36889313}, 
pages = {1391--1401.e5}, 
number = {9}, 
volume = {111}
}

@article{Fries2005, 
year = {2005}, 
title = {{A mechanism for cognitive dynamics: neuronal communication through neuronal coherence}}, 
author = {Fries, Pascal}, 
journal = {Trends in cognitive sciences}, 
issn = {1364-6613}, 
doi = {10.1016/j.tics.2005.08.011}, 
pmid = {16150631}, 
pages = {474 -- 480}, 
number = {10}, 
volume = {9}, 
language = {eng}, 
month = {10}
}

@article{Fries2015, 
year = {2015}, 
title = {{Rhythms for Cognition: Communication through Coherence.}}, 
author = {Fries, Pascal}, 
journal = {Neuron}, 
issn = {0896-6273}, 
doi = {10.1016/j.neuron.2015.09.034}, 
pmid = {26447583}, 
pmcid = {PMC4605134}, 
pages = {220 235}, 
number = {1}, 
volume = {88}, 
language = {English}, 
month = {10}
}

@article{Palmigiano2017, 
year = {2017}, 
title = {{Flexible information routing by transient synchrony.}}, 
author = {Palmigiano, Agostina and Geisel, Theo and Wolf, Fred and Battaglia, Demian}, 
journal = {Nature Neuroscience}, 
issn = {1097-6256}, 
doi = {10.1038/nn.4569}, 
pmid = {28530664}, 
pages = {1014 1022}, 
number = {7}, 
volume = {20}, 
language = {English}
}

@article{Attwell2001, 
year = {2001}, 
month = {7}, 
title = {{An Energy Budget for Signaling in the Grey Matter of the Brain}}, 
author = {Attwell, David and Laughlin, Simon B.}, 
journal = {Journal of Cerebral Blood Flow \& Metabolism}, 
issn = {0271-678X}, 
doi = {10.1097/00004647-200110000-00001}, 
pmid = {11598490}, 
pages = {1133--1145}, 
number = {10}, 
volume = {21}
}

@article{Howarth2012, 
year = {2012}, 
month = {2}, 
title = {{Updated Energy Budgets for Neural Computation in the Neocortex and Cerebellum}}, 
author = {Howarth, Clare and Gleeson, Padraig and Attwell, David}, 
journal = {Journal of Cerebral Blood Flow \& Metabolism}, 
issn = {0271-678X}, 
doi = {10.1038/jcbfm.2012.35}, 
pmid = {22434069}, 
pmcid = {PMC3390818}, 
pages = {1222--1232}, 
number = {7}, 
volume = {32}
}

@article{Hallermann2012, 
year = {2012}, 
month = {7}, 
title = {{State and location dependence of action potential metabolic cost in cortical pyramidal neurons}}, 
author = {Hallermann, Stefan and Kock, Christiaan P J de and Stuart, Greg J and Kole, Maarten H P}, 
journal = {Nature Neuroscience}, 
issn = {1097-6256}, 
doi = {10.1038/nn.3132}, 
pmid = {22660478}, 
pages = {1007--1014}, 
number = {7}, 
volume = {15}
}

@article{Buzsaki2007, 
year = {2007}, 
month = {12}, 
title = {{Inhibition and Brain Work}}, 
author = {Buzsáki, György and Kaila, Kai and Raichle, Marcus}, 
journal = {Neuron}, 
issn = {0896-6273}, 
doi = {10.1016/j.neuron.2007.11.008}, 
pmid = {18054855}, 
pmcid = {PMC2266612}, 
pages = {771--783}, 
number = {5}, 
volume = {56}
}

@article{Kann2012, 
year = {2012}, 
month = {1}, 
title = {{The Energy Demand of Fast Neuronal Network Oscillations: Insights from Brain Slice Preparations}}, 
author = {Kann, Oliver}, 
journal = {Frontiers in Pharmacology}, 
doi = {10.3389/fphar.2011.00090}, 
pmid = {22291647}, 
pmcid = {PMC3254178}, 
pages = {90}, 
volume = {2}
}

@article{Zhang2010, 
year = {2010}, 
month = {1}, 
title = {{Disease and the brain's dark energy}}, 
author = {Zhang, Dongyang and Raichle, Marcus E.}, 
journal = {Nature Reviews Neurology}, 
issn = {1759-4758}, 
doi = {10.1038/nrneurol.2009.198}, 
pmid = {20057496}, 
pages = {15--28}, 
number = {1}, 
volume = {6}
}

@article{Bernacchia2011, 
year = {2011}, 
title = {{A reservoir of time constants for memory traces in cortical neurons}}, 
author = {Bernacchia, Alberto and Seo, Hyojung and Lee, Daeyeol and Wang, Xiao-Jing}, 
journal = {Nature Neuroscience}, 
issn = {1546-1726}, 
doi = {10.1038/nn.2752}, 
pmid = {21317906}, 
pmcid = {PMC3079398}, 
pages = {366 -- 372}, 
number = {3}, 
volume = {14}, 
language = {eng}, 
month = {03}
}

@incollection{Nicosia2013,
  author    = {Nicosia, Vincenzo and Tang, Jonathan and Mascolo, Cecilia and Musolesi, Mirco and Russo, Giovanni and Latora, Vito},
  title     = {Graph Metrics for Temporal Networks},
  editor    = {Holme, Petter and Saram{\"a}ki, Jari},
  booktitle = {Temporal Networks},
  publisher = {Springer},
  address   = {Berlin, Heidelberg},
  year      = {2013},
  pages     = {15--40},
  doi       = {10.1007/978-3-642-36461-7_2}
}

@article{Cui2013, 
year = {2013}, 
month = {5}, 
title = {{On the clustering coefficients of temporal networks and epidemic dynamics}}, 
author = {Cui, Jing and Zhang, Yi-Qing and Li, Xiang}, 
journal = {2013 IEEE International Symposium on Circuits and Systems (ISCAS)}, 
issn = {0271-4302}, 
doi = {10.1109/ISCAS.2013.6572337}, 
pages = {2299--2302}
}

@article{Ciaperoni2020, 
year = {2020}, 
month = {7}, 
title = {{Relevance of temporal cores for epidemic spread in temporal networks}}, 
author = {Ciaperoni, Martino and Galimberti, Edoardo and Bonchi, Francesco and Cattuto, Ciro and Gullo, Francesco and Barrat, Alain}, 
journal = {Scientific Reports}, 
doi = {10.1038/s41598-020-69464-3}, 
pmid = {32719352}, 
pmcid = {PMC7385111}, 
eprint = {2003.09377}, 
pages = {12529}, 
number = {1}, 
volume = {10}
}

@article{Nematzadeh2014, 
year = {2014}, 
month = {8}, 
title = {{Optimal Network Modularity for Information Diffusion}}, 
author = {Nematzadeh, Azadeh and Ferrara, Emilio and Flammini, Alessandro and Ahn, Yong-Yeol}, 
journal = {Physical Review Letters}, 
issn = {0031-9007}, 
doi = {10.1103/physrevlett.113.088701}, 
pmid = {25192129}, 
eprint = {1401.1257}, 
pages = {088701}, 
number = {8}, 
volume = {113}
}

@article{Mitra2015, 
year = {2015}, 
month = {4}, 
title = {{Lag threads organize the brain’s intrinsic activity}}, 
author = {Mitra, Anish and Snyder, Abraham Z. and Blazey, Tyler and Raichle, Marcus E.}, 
journal = {Proceedings of the National Academy of Sciences}, 
issn = {0027-8424}, 
doi = {10.1073/pnas.1503960112}, 
pmid = {25825720}, 
pmcid = {PMC4418865}, 
pages = {E2235--E2244}, 
number = {17}, 
volume = {112}
}

@article{Liu2018, 
year = {2018}, 
month = {10}, 
title = {{Co-activation patterns in resting-state fMRI signals}}, 
author = {Liu, Xiao and Zhang, Nanyin and Chang, Catie and Duyn, Jeff H.}, 
journal = {NeuroImage}, 
issn = {1053-8119}, 
doi = {10.1016/j.neuroimage.2018.01.041}, 
pmid = {29355767}, 
pmcid = {PMC6082734}, 
pages = {485--494}, 
number = {Pt B}, 
volume = {180}
}

@article{Tagliazucchi2012, 
year = {2012}, 
month = {2}, 
title = {{Criticality in Large-Scale Brain fMRI Dynamics Unveiled by a Novel Point Process Analysis}}, 
author = {Tagliazucchi, Enzo and Balenzuela, Pablo and Fraiman, Daniel and Chialvo, Dante R.}, 
journal = {Frontiers in Physiology}, 
doi = {10.3389/fphys.2012.00015}, 
pmid = {22347863}, 
pmcid = {PMC3274757}, 
pages = {15}, 
volume = {3}
}

@article{Hengen2025, 
year = {2025}, 
month = {8}, 
title = {{Is criticality a unified setpoint of brain function?}}, 
author = {Hengen, Keith B. and Shew, Woodrow L.}, 
journal = {Neuron}, 
issn = {0896-6273}, 
doi = {10.1016/j.neuron.2025.05.020}, 
pmid = {40555236}, 
pmcid = {PMC12374783}, 
pages = {2582--2598.e2}, 
number = {16}, 
volume = {113}
}

@article{Newman2007, 
year = {2007}, 
month = {1}, 
title = {{Component sizes in networks with arbitrary degree distributions}}, 
author = {Newman, M. E. J.}, 
journal = {Physical Review E}, 
issn = {1539-3755}, 
doi = {10.1103/PhysRevE.76.045101}, 
pmid = {17995046}, 
eprint = {0707.0080}, 
pages = {045101}, 
number = {4}, 
volume = {76}
}

@article{Cirigliano2024, 
year = {2024}, 
month = {12}, 
title = {{Scaling and universality for percolation in random networks: A unified view}}, 
author = {Cirigliano, Lorenzo and Timár, Gábor and Castellano, Claudio}, 
journal = {Physical Review E}, 
issn = {2470-0045}, 
doi = {10.1103/PhysRevE.110.064303}, 
pmid = {39916202}, 
eprint = {2408.05125}, 
pages = {064303}, 
number = {6}, 
volume = {110}
}

@article{Krings2012, 
year = {2012}, 
month = {5}, 
title = {{Effects of time window size and placement on the structure of an aggregated communication network}}, 
author = {Krings, Gautier and Karsai, Márton and Bernhardsson, Sebastian and Blondel, Vincent D and Saramäki, Jari}, 
journal = {EPJ Data Science}, 
doi = {10.1140/epjds4}, 
pages = {4}, 
number = {1}, 
volume = {1}
}

@article{Shew2011, 
year = {2011}, 
title = {{Information capacity and transmission are maximized in balanced cortical networks with neuronal avalanches.}}, 
author = {Shew, Woodrow L and Yang, Hongdian and Yu, Shan and Roy, Rajarshi and Plenz, Dietmar}, 
journal = {Journal of Neuroscience}, 
issn = {0270-6474}, 
doi = {10.1523/jneurosci.4637-10.2011}, 
pmid = {21209189}, 
pmcid = {PMC3082868}, 
pages = {55 -- 63}, 
number = {1}, 
volume = {31}, 
language = {English}, 
month = {01}
}

@article{Faskowitz2019, 
year = {2019}, 
title = {{Edge-centric functional network representations of human cerebral cortex reveal overlapping system-level architecture.}}, 
author = {Faskowitz, Joshua and Esfahlani, Farnaz Zamani and Jo, Youngheun and Sporns, Olaf and Betzel, Richard F}, 
journal = {Nature neuroscience}, 
doi = {10.1038/s41593-020-00719-y}, 
pmid = {33077948}, 
pages = {1644--1654}, 
number = {12}, 
volume = {23}
}

@article{Cole2014, 
year = {2014}, 
title = {{Intrinsic and task-evoked network architectures of the human brain.}}, 
author = {Cole, Michael W and Bassett, Danielle S and Power, Jonathan D and Braver, Todd S and Petersen, Steven E}, 
journal = {Neuron}, 
issn = {0896-6273}, 
doi = {10.1016/j.neuron.2014.05.014}, 
pmid = {24991964}, 
pmcid = {PMC4082806}, 
pages = {238 251}, 
number = {1}, 
volume = {83}, 
language = {English}, 
month = {07}
}

@article{VandeVille2010, 
year = {2010}, 
title = {{EEG microstate sequences in healthy humans at rest reveal scale-free dynamics.}}, 
author = {Van de Ville, Dimitri and Britz, Juliane and Michel, Christoph M}, 
journal = {Proceedings of the National Academy of Sciences of the United States of America}, 
issn = {0027-8424}, 
doi = {10.1073/pnas.1007841107}, 
pmid = {20921381}, 
pmcid = {PMC2964192}, 
pages = {18179 -- 18184}, 
number = {42}, 
volume = {107}, 
language = {English}, 
month = {10}
}

@article{Glomb2018, 
year = {2018}, 
title = {{Stereotypical modulations in dynamic functional connectivity explained by changes in BOLD variance}}, 
author = {Glomb, Katharina and Ponce-Alvarez, Adrián and Gilson, Matthieu and Ritter, Petra and Deco, Gustavo}, 
journal = {NeuroImage}, 
issn = {1053-8119}, 
doi = {10.1016/j.neuroimage.2017.12.074}, 
pmid = {29294385}, 
pages = {40--54}, 
number = {25}, 
volume = {171}
}

@article{termenon2016reliability,
  title={Reliability of graph analysis of resting state fMRI using test-retest dataset from the Human Connectome Project},
  author={Termenon, Ma{\"\i}t{\'e} and Jaillard, Assia and Delon-Martin, Chantal and Achard, Sophie},
  journal={Neuroimage},
  volume={142},
  pages={172--187},
  year={2016},
  publisher={Elsevier}
}

@article{tzourio2002automated,
  title={Automated anatomical labeling of activations in SPM using a macroscopic anatomical parcellation of the MNI MRI single-subject brain},
  author={Tzourio-Mazoyer, Nathalie and Landeau, Brigitte and Papathanassiou, Dimitri and Crivello, Fabrice and Etard, Octave and Delcroix, Nicolas and Mazoyer, Bernard and Joliot, Marc},
  journal={Neuroimage},
  volume={15},
  number={1},
  pages={273--289},
  year={2002},
  publisher={Elsevier}
}

@article{mengiste2015effect,
  title={Effect of edge pruning on structural controllability and observability of complex networks},
  author={Mengiste, Simachew Abebe and Aertsen, Ad and Kumar, Arvind},
  journal={Scientific reports},
  volume={5},
  number={1},
  pages={18145},
  year={2015},
  publisher={Nature Publishing Group UK London}
}

\newpage
\section*{Tables}

\begin{table}[h]
\centering
\caption{Temporal-network measures used in the analysis, grouped by functional role: integration, segregation, memory, and statism/dynamism.}
\begin{tabular}{p{0.28\linewidth}p{0.64\linewidth}}
\toprule
\textbf{Measure} & \textbf{Interpretation} \\
\midrule

\multicolumn{2}{l}{\emph{Integration}} \\
&\\
\textbf{Smart irrigation} & Number of reachable node pairs under earliest-arrival. \\
\textbf{Smart latency} & Mean earliest-arrival time among reachable node pairs. \\
\textbf{Smart resistance} & Penalized earliest-arrival time incorporating unreachable pairs. \\
\textbf{Smart inaccessibility} & Number of unreachable node pairs under earliest-arrival scheme. \\

\textbf{Drunk irrigation} & Number of reachable node pairs under random-walk propagation. \\
\textbf{Drunk latency} & Mean first-passage time under random-walk propagation. \\
\textbf{Drunk resistance} & Penalized first-passage time including unreachable pairs. \\
\textbf{Drunk inaccessibility} & Number of unreachable node pairs under random-walk scheme. \\

\midrule
\multicolumn{2}{l}{\emph{Segregation}} \\
&\\
\textbf{Return latency} &  Earliest time for a node to return to itself via a time-respecting path. \\
\textbf{Retention} & Fraction of current neighbors that have been encountered previously (temporal partner re-use). \\
\textbf{Static clustering} & Tendency of node neighborhoods to be themselves interconnected per frame. \\
\textbf{Temporal clustering} & Tendency of triangle closure across successive frames within a short temporal window ($\Delta=1$). \\

\midrule
\multicolumn{2}{l}{\emph{Statism/dynamism}} \\
&\\
\textbf{Similarity} & Cosine similarity between consecutive frames. \\
\textbf{Net fluidity} & Combination of global edge-activation entropy and low frame similarity, capturing temporal reconfiguration. \\

\bottomrule
\end{tabular}
\label{tab:temporal_measures}
\end{table}

\newpage

\section*{Extended Data - Figures}
\setcounter{figure}{0}
\renewcommand{\thefigure}{S\arabic{figure}}

\begin{figure}[H]
\centering
\includegraphics[width=\textwidth]{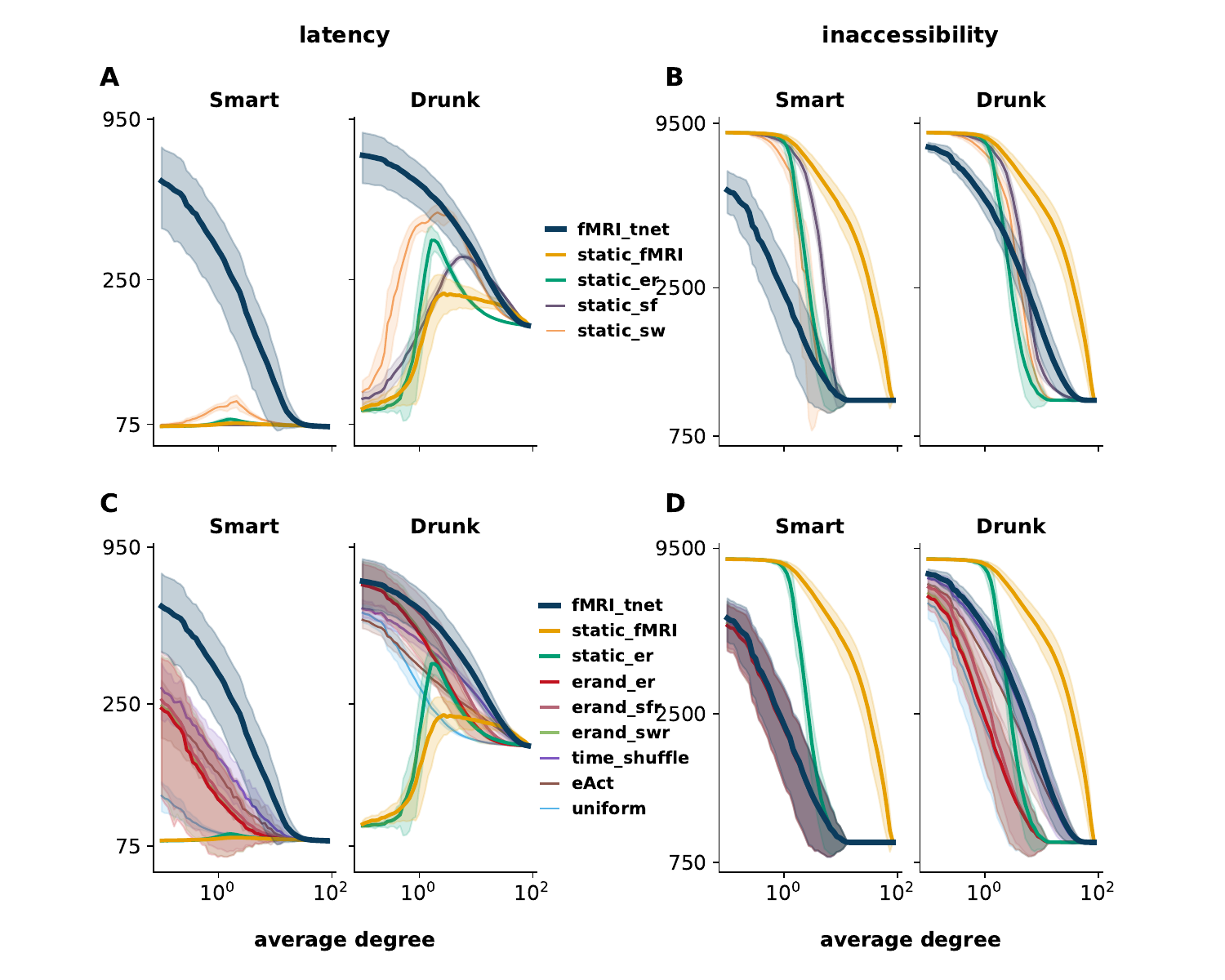}
\caption{\textbf{The lower irrigation resistance does not reflect faster realized diffusion, but fewer disconnected source-target pairs.} A, Latency computed only over source-target pairs that successfully communicate within the observation horizon (i.e. ignoring targets for which information dispatch failed). Under this definition, temporal reconfiguration does not speed up diffusion and is often slower than static controls, consistent with the waiting-time effect induced by intermittent link availability. B, Number of communicating source-target pairs (or, equivalently, fraction of reachable pairs) within the same observation horizon. Temporal networks strongly increase the number of successful communications relative to static controls in the sparse regime. Thus, the reduction in irrigation resistance reported in Fig.~\ref{fig2} does not arise because successful paths are intrinsically faster in dynamic networks, but because dynamicity converts many source-target pairs that would remain disconnected in static networks into reachable ones, thereby removing large numbers of $T+1$ penalty contributions. All model labels are as in Fig.~\ref{fig2}.} \label{SupFig1}
\end{figure}

\begin{figure}[H]
\centering
\includegraphics[width=\textwidth]{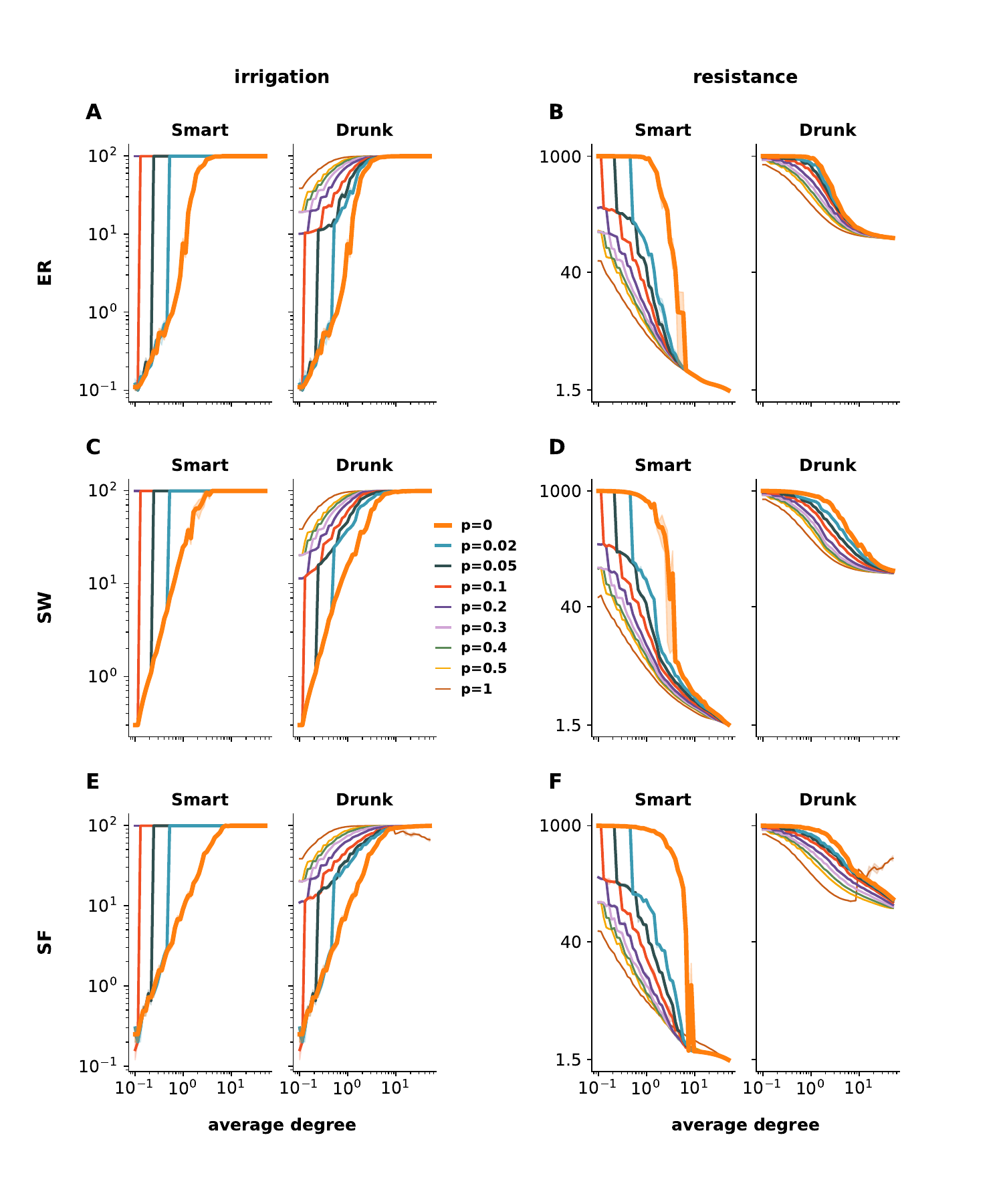}
\caption{\textbf{Communication gains increase monotonically with temporal randomization in null temporal networks.} A, Irrigation as a function of average degree for Erdős--Rényi (ER) temporal networks with progressively increasing rewiring probability, spanning the continuum from static networks to highly reconfiguring temporal graphs. B, Irrigation resistance for the same ER ensembles. Increasing temporal randomization progressively improves information dispatch, with the strongest benefit in the sparse, high-cost regime. This shows that the communication advantage of dynamicity arises generically from redistributing a fixed link budget across time, rather than from specific empirical features of resting-state dFC. The same analysis is shown for progressively edge-randomized small-world (SW) and scale-free (SF) ensembles: irrigation and irrigation resistance for SW networks in C and D, respectively, and for SF networks in E and F.}\label{SupFig2}
\end{figure}

\begin{figure}[H]
\centering
\includegraphics[width=\textwidth]{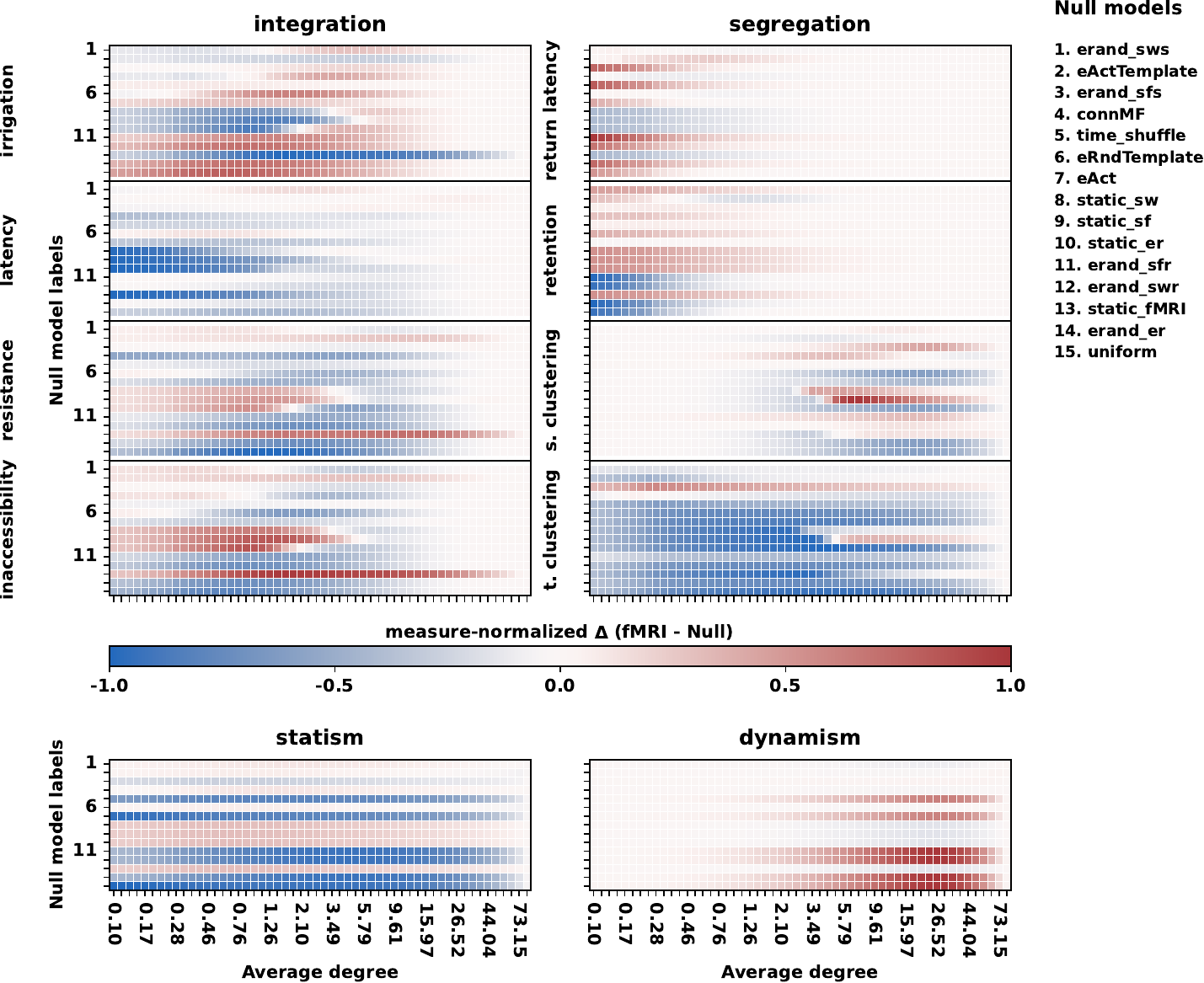}
\caption{\textbf{Synoptic comparison of null-model deviations from empirical temporal functional connectivity across integration, segregation and temporal-organization metrics.} Heat maps show the measure-normalized $\Delta$ between each null model and the empirical resting-state temporal network (\texttt{fMRI tnet}) as a function of average degree. Rows correspond to null models, numbered as in the legend and labeled using the abbreviations introduced in Figs.~\ref{fig2} and~\ref{fig4}. The upper panels summarize measures related to integration (left: irrigation, latency, resistance and inaccessibility) and segregation (right: return latency, retention, spatial clustering and temporal clustering). The lower panels summarize temporal-organization metrics, with ``statism'' quantified by cosine similarity between consecutive frames (bottom left) and ``dynamism'' quantified by net fluidity (bottom right). Blue values indicate regimes in which the null model exceeds the empirical \texttt{fMRI tnet} for the corresponding measure, whereas red values indicate regimes in which the empirical temporal network shows larger values. This representation provides a compact overview of how each null model departs from empirical dFC across metrics and cost regimes.}\label{SupFig3}
\end{figure}

\newpage
\section*{Extended Data - Tables}
\setcounter{table}{0}
\renewcommand{\thetable}{S\arabic{table}}

\begin{table}[h]
\centering
\caption{Brain regions used in the analysis based on the Automated Anatomical Labeling (AAL) atlas. Regions are grouped by anatomical lobe.}
\begin{tabular}{p{0.28\linewidth}p{0.64\linewidth}}
\toprule
\textbf{Lobe / System} & \textbf{Regions (bilateral unless noted)} \\
\midrule

Frontal &
Precentral gyrus, superior frontal gyrus (dorsal, medial, orbital), middle frontal gyrus, inferior frontal gyrus (opercular, triangular, orbital), supplementary motor area, olfactory cortex, rectus \\

Parietal &
Postcentral gyrus, superior parietal gyrus, inferior parietal gyrus (supramarginal, angular), precuneus \\

Temporal &
Superior temporal gyrus, middle temporal gyrus, inferior temporal gyrus, Heschl gyrus, temporal pole (superior and middle) \\

Occipital &
Superior occipital gyrus, middle occipital gyrus, inferior occipital gyrus, calcarine cortex, cuneus, lingual gyrus, fusiform gyrus \\

Limbic &
Hippocampus, parahippocampal gyrus, amygdala, anterior cingulate cortex, posterior cingulate cortex \\

Subcortical &
Caudate nucleus, putamen, pallidum, thalamus \\

Insular &
Insula \\

\bottomrule
\end{tabular}
\label{tab:aal_regions}
\end{table}

\end{document}